\documentclass[onecolumn]{mn2e}
\newif\ifAMStwofonts

\def\pmb#1{\mbox{\boldmath$#1$}}
\def\gtsim {>\kern-1.2em\lower1.1ex\hbox{$\sim$}}
\def\ltsim {<\kern-1.2em\lower1.1ex\hbox{$\sim$}}
\def\gtsim {>\kern-1.2em\lower1.1ex\hbox{$\sim$}}
\def\ltsim {<\kern-1.2em\lower1.1ex\hbox{$\sim$}}
\def\ref{\hangindent=1pc \hangafter=1 \noindent}
\usepackage{epsfig}
\begin{document}

\title[Light curves of oscillating neutron stars]
{Light curves of oscillating neutron stars}

\author[U. Lee \& T.E. Strohmayer]{Umin Lee$^1$ \thanks{E-mail:
lee@astr.tohoku.ac.jp; stroh@clarence.gsfc.nasa.gov} and Tod
E. Strohmayer$^2$ \\$^1$Astronomical Institute, Tohoku University,
Sendai, Miyagi 980-8578, Japan \\$^2$Laboratory for High Energy
Astrophysics, NASA's Goddard Space Flight Center, Greenbelt, MD 20771,
USA}

\date{Typeset \today ; Received / Accepted}
\maketitle


\begin{abstract} 
We calculate light curves produced by $r$-modes with small azimuthal
wavenumbers, $m$, propagating in the surface fluid ocean of rotating
neutron stars. We include relativistic effects due to rapid rotation,
and propagate photons from the stellar surface to a distant observer
using the Schwarzschild metric. The wave motions of the surface
$r$-modes are confined to the equatorial region of the star, and the
surface pattern of the temperature variation can be either symmetric
(for even modes) or anti-symmetric (for odd modes) with respect to the
equator. Since for the surface $r$-modes the oscillation frequency in
the corotating frame of the star is much smaller than the rotation
frequency, $\Omega$, we employ the approximation in which the
oscillation frequency in the inertial frame, $\sigma$, is given by
$\sigma= -m\Omega$.  We find that the $even$, $m = 1$ $r$-mode
produces the largest light variations. The dominant Fourier component
in the light curves of these modes is the fundamental having
$\sigma=-\Omega$, and the first harmonic component having
$\sigma=-2\Omega$ is always negligible in comparison.  The dominant
Fourier component of the even, $m=2$ $r$-modes is the first harmonic.
Although the $odd$ $r$-modes produce smaller amplitude light
variations compared with the $even$ modes, the light curves of the
former have a stronger first harmonic component.  If both $m=1$ and
$2$ $r$-modes are excited simultaneously, a rich variety of light
curves is possible, including those having an appreciable first
harmonic component. We show that the phase difference,
$\delta-\delta_E$, between the bolometric light curve and that at a
particular photon energy can possibly be used as a probe of the
stellar compactness, $R/M$, where $R$ and $M$ are the radius and mass
of the star.
\end{abstract}



\section{Introduction}

It is now widely accepted that millisecond oscillations during
thermonuclear X-ray bursts (hereafter, burst oscillations) on
accreting, weakly magnetized neutron stars in low mass X-ray binary
(LMXB) systems are produced by spin modulation of a slowly moving,
non-uniform brightness pattern on the stellar surface (see Strohmayer
et al. 1996; Strohmayer \& Bildsten 2004, van der Klis 2004). In
addition to revealing the spin frequency these oscillations encode
information about the global properties of the neutron star. A number
of attempts have been made to model burst oscillations and thus to
infer these properties (e.g., Nath, Strohmayer \& Swank 2001; Muno,
\"Ozel \& Chakrabarty 2002; Poutanen \& Gierli\'nski 2003,
Bhattacharyya et al. 2005), however, uncertainties associated with the
exact nature of the brightness pattern makes a unique interpretation
of the data problematic.

As of February, 2005, the number of LMXBs that exhibit burst
oscillations amounts to 13, and two of these contain millisecond X-ray
pulsars (see, e.g., Chakrabarty et al. 2003; Strohmayer et al. 2003).
For most burst oscillations a single oscillation frequency is found
that displays a positive frequency drift $\Delta\nu_{\rm burst}$ of a
few Hz from the start to the tail of a burst. In the decaying tail the
frequency often approaches an asymptotic value (e.g., Strohmayer \&
Markwardt 1999).  The asymptotic frequencies are stable to a few parts
in 1000 in bursts observed over several years from any given source
(Strohmayer et al. 1998b; Muno et al. 2002a).  Background subtracted
light curves for the burst oscillations are in most cases well fitted
by a single sinusoid, the frequency of which we will call the
fundamental frequency. Harmonic content in the light curves is
generally small (Strohmayer \& Markwardt 1999; Muno et al.
2002b). Indeed, to date there exists only one strong detection of a
first harmonic, in the millisecond X-ray pulsar XTE J1814-338
(Strohmayer et al. 2003, Bhattacharyya et al. 2005).  On average the rms
amplitudes of the burst oscillations are 2\%-10\%, and generally
increase with photon energy (Muno et al. 2003). Muno et al.(2003) also
explored the energy dependence of the oscillation phases of burst
oscillations. They found either constant phase lags with energy or in
some cases evidence for modest hard lags, behavior that is opposite to
what one would expect from relativistic motion of a simple hot spot on
the stellar surface (e.g., Ford 1999).  Interestingly, where phase
lags have been measured in the accreting millisecond pulsars they all
show soft lags, consistent with a rotating hot spot model (Cui, Morgan
\& Titarchuk 1998; Poutanen \& Gierli\'nski 2003).

As noted above, a non-uniform brightness distribution on the surface
of rotating neutron stars is almost certainly responsible for the
burst oscillations, and several models have so far been proposed to
explain the inhomogeneous pattern.  As a typical model, Strohmayer et
al. (1997) suggested the oscillation originates from hot regions (hot
spots) in a thermonuclear burning layer that expands and decouples
from the neutron star at the start of the burst.  Although a simple
hot spot appears to be consistent with oscillations observed during
the onset of some bursts (see Strohmayer et al. 1998a), further
calculations suggested that a rigidly rotating, hydrostatically
expanding burning layer produces too small a frequency drift (Cumming
\& Bildsten 2000, Cumming et al. 2002).

There have been numerous efforts to calculate light curves produced by
hot spots on the surface of rotating neutron stars (e.g., Pechenick et
al. 1983, Chen \& Shaham 1989, Strohmayer 1992, Page 1995, Braje et
al. 2000, Weinberg et al. 2001, Muno et al.  2002, Nath et al. 2002,
Poutanen \& Gierli\'nski 2003, Bhattacharyya et al. 2005). In the
modeling of the light curves many parameters are employed, for
example, the number of hot spots, their location with respect to the
rotation axis, their angular sizes, the inclination angle between the
line of sight and the rotation axis, the angular dependence of the
specific intensity, the temperature of the spots, the temperature
contrast between the spots and their surroundings, the mass, radius,
and spin frequency of the star.  As noted above, burst oscillation
light curves are remarkably sinusoidal.  Although the light curve
amplitudes are dependent on the relativistic parameter $R/M$, which is
a parameter of particular importance, it is obvious that there remain
parameters other than $R/M$ that affect the amplitudes, such as the
angular sizes and locations of the hot spots and the inclination angle
(e.g., Pechenick et al. 1983, Weinberg et al. 2001, Muno et al. 2002b).
In this sense it is quite important to find other observational
quantities that can be used to obtain independent constraints on these
parameters. The energy dependencies of phase lags are an example of
potentially important observational quantities that carry useful
information about the neutron star.  Taking account of the special
relativistic Doppler boosting of X-ray photons due to rapid rotation
of the star (Ford 1999, Weinberg et al. 2001), the hot spot model can
produce hard leads more or less consistent with those observed in the
millisecond X-ray pulsar SAX J1808.4-3658 and IGR J00291+5934
(Poutanen \& Gierli\'nski 2003; Galloway et al. 2005), but the model
appears to be at odds with the suggestion of hard lags in some burst
oscillations (Muno et al. 2003).  Although it may be possible to
produce hard lags via a hot Comptonizing corona, it is worth exploring
in more detail whether more complex brightness patterns (such as
oscillation mode patterns) on the stellar surface might be consistent
with the production of hard lags.

By inferring the hot spot size using the observationally obtained
blackbody flux $F_{bb}$, Poutanen \& Gierli\'nski (2003) have
succeeded in estimating the radius of the millisecond X-ray pulsar SAX
J1808.4-3658 assuming an appropriate range of mass for the star.
Recently, Strohmayer et al. (2003) have found the first harmonic
component in the light curves of burst oscillations from the accreting
millisecond X-ray pulsar XTE J1814-338, and Bhattacharyya et
al. (2005) obtained a useful constraint on the ratio $R/M$ assuming a
single hot spot model.

As alluded to above, the hot spot model is not always successful in
producing the observed light curves of the burst oscillations.  As
discussed by Muno et al. (2002b), a small hot spot in a completely dark
background is likely to produce light curves with too large an
amplitude as well as having strong harmonic components not seen
observationally.  To suppress the large amplitudes and the harmonics
one has to assume, for example, a hot spot of angular size $\alpha\sim
\pi/2$ or almost completely antipodal hot spots, the assumptions of
which are not necessarily physically well motivated. It is therefore
important to pursue alternatives to the hot spot model.

In this paper, we calculate light curves of rotating neutron stars
assuming the surface temperature perturbation pattern caused by
surface $r$-modes propagating in the fluid ocean of a mass accreting
neutron star (Lee 2004). It was Heyl (2004) who first argued that the
$r$-modes with low $m$ may play a possible role for burst
oscillations.  The surface $r$-modes are rotationally induced
oscillations with very low frequencies in the corotating frame of the
star.  The $r$-modes have dominant vorticity
$(\nabla\times\pmb{v}^\prime)_r$ and are basically just large scale
vortexes (see Spitkovsky, Levin, \& Ushomirsky 2002).  If we assume
axisymmetry of the equilibrium configuration of the rotating star, the
time dependence of the temperature perturbation may be given by
$e^{i\sigma t}$, where $\sigma$ denotes the oscillation frequency in
the inertial frame.  If we let $\omega$ and $\Omega$ denote the
oscillation frequency in the corotating frame and the spin frequency
of the star, respectively, we have $\sigma=\omega-m\Omega$, which
leads to $\sigma\approx -m\Omega$ for the surface $r$-modes, since
$|\omega|<<|m\Omega|$, where $m$ is an integer representing the
azimuthal wave number around the rotation axis.  The time dependence
of the perturbation is thus to good approximation given by
$e^{-im\Omega t}$.  Since only the oscillation modes having low $m$
will produce light variations of appreciable amplitude, in this paper,
we only consider the surface $r$-modes with $m=1$ or $m=2$.  We
briefly describe the method of calculation in \S 2. In \S 3 we discuss
some of our numerical results with relevance to burst oscillations. 
\S 4 is for a discussion, and we
give a summary of our principal conclusions in \S 5.

\section{method of solution}

The method of calculating light curves of rotating neutron stars is
similar to those discussed by Pechenick et al. (1983), Page (1995),
and Weinberg et al. (2001). Instead of presuming the presence of a hot
spot on the surface of a rotating neutron star, we assume that the
temperature varies accross the surface with the angular distribution
appropriate for a surface $r$-mode propagating in the fluid ocean of
the star.  The surface $r$-modes, which have very low oscillation
frequencies in the corotating frame of the star, can plausibly be
excited during thermonuclear bursts (Lee 2004).

To describe oscillations of rotating neutron stars, we employ an $xyz$
coordinate system whose origin is at the center of the star with the
rotation axis along the $z$ direction.  For convenience, we also
assume that the observer is in the $x$-$z$ plane.  In this coordinate
system, the oscillations of rotating neutron stars are described by
employing spherical polar coordinates $(r,\theta,\phi)$ with the axis
$\theta=0$ being the $z$-axis.  Then, the displacement vector
$\pmb{\xi}(r,\theta,\phi,t)$ is given by
\begin{equation}
\xi_r=\sum_{j=1}^{\infty}S_{l_j}Y_{l_j}^m e^{i\sigma t},
\end{equation}
\begin{equation}
\xi_\theta=\sum_{j=1}^{\infty}
\left(H_{l_j}{\partial\over\partial\theta}Y_{l_j}^m
+T_{l^\prime_j}{1\over\sin\theta}{\partial\over\partial\phi}Y_{l^\prime_j}^m
\right)
e^{i\sigma t},
\end{equation}
\begin{equation}
\xi_\phi=\sum_{j=1}^{\infty}
\left(H_{l_j}{1\over\sin\theta}{\partial\over\partial\phi}Y_{l_j}^m
-T_{l^\prime_j}{\partial\over\partial\theta}Y_{l^\prime_j}^m\right)
e^{i\sigma t},
\end{equation}
and the Lagrangian perturbation of the temperature is given by
\begin{equation}
\delta T=\sum_{j=1}^{\infty}\delta T_{l_j}Y_{l_j}^me^{i\sigma t},
\end{equation}
where $\sigma$ is the oscillation frequency in the inertial frame, and
$Y_l^m$ denotes the spherical harmonic function with degree $l$ and
azimuthal wave number $m$, and $l_j=|m|+2(j-1)$ and $l^\prime_j=l_j+1$
for even modes, and $l_j=|m|+2(j-1)+1$ and $l^\prime_j=l_j-1$ for odd
modes, and $j=1,~2,~\cdots$.  The functions
$\pmb{\xi}(r,\theta,\phi,t)$ and $\delta T$ are obtained as a solution
to linear differential equations for oscillations of a rotating star,
and we need to specify a normalization condition to determine the
amplitudes.  Since the surface $r$-modes have dominant toroidal
components $T_{l^\prime_j}$, we will employ the amplitude
normalization given by $|{\rm Re}[iT_{l^\prime_1}(R)/R]|=1$.

The surface temperature of a rotating star that pulsates in various
oscillation modes may formally be given by
\begin{equation}
T(R,\theta,\phi,t)=T_0
+{\rm Re}\left\{
\sum_{\alpha}C_{\alpha}
\left[\sum_{j\ge1}\delta T_{l_j}^{\alpha}(R)
Y_{l_j}^m(\theta,0)\right]e^{ i(m\phi+\sigma_{\alpha} t)}\right\},
\end{equation}
where $T_0$ is the mean surface temperature, $\alpha$ indicates a
combination of indices that distinguish various oscillation modes, and
$\sigma_{\alpha}$ and $C_{\alpha}$ denote, respectively, the oscillation
frequency and a complex constant representing the amplitudes of the
oscillation mode $\alpha$.  Note that by introducing $C_\alpha$ in
equation (5) we mean $|{\rm
Re}[iT^{\alpha}_{l^\prime_1}(R)/R]|=|C_\alpha|$.  Using the
oscillation frequency $\omega=\sigma+m\Omega$ in the corotating frame,
we have
\begin{equation}
m\phi+\sigma_{\alpha} t=m\hat\phi+\omega_{\alpha} t,
\end{equation}
where $\hat\phi=\phi-\Omega t$ is the longitude in the corotating
frame.

To consider photon trajectories around a neutron star, it is convenient
to introduce another coordinate system $x^\prime y^\prime z^\prime$,
for which the origin is at the center of the star, and the $z^\prime$
axis is pointing to the observer and $y=y^\prime$.  If we let $i$
denote the inclination angle between the $z$ axis and the $z^\prime$
axis, we have
\begin{equation}
x=x^\prime\cos i+z^\prime\sin i, \quad\quad y=y^\prime, \quad\quad
z=-x^\prime\sin i +z^\prime\cos i,
\end{equation}
and it is easy to derive the relation between the two spherical polar
coordinate systems $(r,\theta,\phi)$ and
$(r^\prime,\theta^\prime,\phi^\prime)$, where the axis of
$\theta^\prime=0$ is the $z^\prime$ axis.

If we assume the Schwartzschild metric around the neutron star with
mass $M$ and radius $R$, the number flux, per photon energy
$E_\infty$, of thermal photons emitted from the stellar surface and
received by a distant observer may be given by (e.g., Page 1995)
\begin{equation}
{1\over S}{d^2N(E_\infty)\over dt_\infty dE_\infty}
={2\pi\over c^2h^3}{R^2_\infty\over D^2}E^2_\infty
\int_0^12qdq\int_0^{2\pi}{d\phi^\prime\over 2\pi}
{1\over e^{f e^{-\eta}E_\infty/k_BT}-1}
\equiv {2\pi\over c^2h^3}{R^2_\infty\over D^2}
{E^2_\infty \over e^{e^{-\eta}E_\infty/k_BT_0}-1}G_{E_\infty}(t_\infty),
\end{equation}
and integrating equation (8) with respect to the variable $E_\infty$
leads to
\begin{equation}
{1\over S}{dN(E_\infty)\over dt_\infty}=2.404\times{2\pi\over
c^2h^3}{R^2_\infty\over D^2}
\int_0^12qdq\int_0^{2\pi}{d\phi^\prime\over 2\pi} \left({k_BT\over f
e^{-\eta}}\right)^3 \equiv {4.808\pi\over c^2h^3}{R^2_\infty\over D^2}
\left({k_BT_0\over e^{-\eta}}\right)^3G(t_\infty),
\end{equation}
where $D$ is the distance to the observer, $T$ is the surface
temperature, $k_B$ is the Boltzmann constant, $c$ is the velocity of
light, and $h$ is the Planck constant, and we have neglected, for
simplicity, the interstellar absorption effect and the dependence of
the effective area $S$ of the detector on the photon energy. Here,
\begin{equation}
R_\infty=e^{-\eta} R
\end{equation}
is the radius of the star seen by a distant observer, and
$dt_\infty=e^{-\eta}dt$ denotes the increment of the coordinate time
$t_\infty$, where
\begin{equation}
e^\eta=\sqrt{1-R_G/R}
\end{equation}
and $R_G=2GM/c^2$ is the Schwarzschild radius, and $dt$ is the
increment of the proper time at the surface of the star.  Note that
$\Omega_\infty=e^\eta\Omega$ since $\Omega_\infty dt_\infty=\Omega
dt$, where $\Omega_\infty$ is the angular spin frequency of the star
seen from infinity.  If we let $E_e$ and $E_\infty$ denote,
respectively, the photon energy in the corotating frame of the star at
the surface and the photon energy recieved by a distant observer, we
have
\begin{equation}
E_e=f e^{-\eta} E_\infty,
\end{equation}
where the factor $f$, representing the Doppler shift in the photon
energy from $ e^{-\eta} E_\infty$ in the non-rotating frame at the
stellar surface to $E_e$ in the co-rotating frame, is given by
\begin{equation}
f=\gamma  \left(1-\pmb{v}\cdot\pmb{o}/c\right),
\end{equation}
where
$\pmb{v}=\pmb{\Omega}\times\pmb{R}=R\Omega\sin\theta\pmb{e}_\phi$ with
$\pmb{e}_\phi$ being the unit vector in the azimuthal direction and
$\pmb{o}$ is the unit vector along the trajectory of a photon at the
surface, and $\gamma=1/\sqrt{1-\pmb{v}^2/c^2}$.  Note that the four
vector of a photon in the non-rotating frame may be given by
$o^\mu=e^{-\eta}E_\infty(1,\pmb{o})$.  Using the angle
$\zeta(\theta^\prime)$ between the local normal and the photon
trajectory, at the surface, reaching the observer at infinity, the
variable $q$ in equations (8) and (9) is defined by
\begin{equation}
q\equiv\sin\zeta,
\end{equation}
and the relation between $\theta^\prime$ and $q$ 
is determined by the integration given by
\begin{equation}
\theta^\prime(q)=\int_0^{R_G/2R}
{qdu\over\sqrt{\left(1-{R_G/R}\right){R_G/2R}
-(1-2u)u^2q^2}}.
\end{equation}
The maximum value $\theta^\prime_{\max}$ occurs when $q=1$ and is shown
as a function of the radius $R$ in Figure 1 for $M=1.4M_\odot$.
Since $\pmb{o}$ at the stellar surface is given by
\begin{equation}
\pmb{o}=\sin(\theta^\prime-\zeta)\cos\phi^\prime\pmb{i}^\prime
+\sin(\theta^\prime-\zeta)\sin\phi^\prime\pmb{j}^\prime
+\cos(\theta^\prime-\zeta)\pmb{k}^\prime,
\end{equation}
where $\pmb{i}^\prime$, $\pmb{j}^\prime$, and $\pmb{k}^\prime$ are the
orthonormal vectors parallel to the $x^\prime$, $y^\prime$, and
$z^\prime$ axes, respectively, we have
\begin{equation}
{\pmb{v}\cdot\pmb{o}\over\Omega R\sin\theta}=\sin(\theta^\prime-\zeta)
\left(\cos\phi\sin\phi^\prime-\sin\phi\cos\phi^\prime\cos i\right)
-\cos(\theta^\prime-\zeta)\sin\phi\sin i.
\end{equation}
Note that without light bending so that $\theta^\prime=\zeta$, we simply have 
$\pmb{v}\cdot\pmb{o}=-\Omega R\sin\phi\sin i\sin\theta$.
The aberration effect may be given by
\begin{equation}
\cos\zeta_e={e^{-\eta}E_\infty\over E_e}\cos\zeta,
\end{equation}
where $\zeta_e$ is the angle measured in the corotating frame of the star.

For waves whose $\phi$ and $t$ dependence is given by
$e^{im\phi+i\sigma t}$, it is easy to incorporate the time-delay
effect that arises because of the difference in the path lengths of
photons emitted from different points on the surface of the star.
Since the travel time of a photon from a point having the impact
parameter $b=qR_\infty$ at the surface of the star to an observer at
infinity is given by
\begin{equation}
t_\infty(b)={R_G\over c}\int_{R}^\infty{1\over\sqrt{1-(1-R_G/r)(b/r)^2}}
\left(1-{R_G\over r}\right)^{-1}{dr\over R_G},
\end{equation}
the time-lag may be defined by
\begin{equation}
\Delta t_\infty(b)\equiv t_\infty(b)- t_\infty(0).
\end{equation}
Replacing $t_\infty$ by $t_\infty-\Delta t_\infty(b)$, we rewrite
the time and $\phi$ dependence $e^{im\phi+i\sigma t}$ of the perturbations as
\begin{equation}
e^{im\phi+i\sigma t}e^{-ie^\eta\sigma\Delta t_\infty(b)}
\end{equation}
so that we can take account of the time-lag effects in light curve
calculations (see e.g., Cadeau et al. 2005).

For the light curves $G(t)$ or $G_{E_\infty}(t)$, we calculate the discrete
Fourier transform $a_j$ ($j=-N/2,~\cdots,N/2-1$) defined by
\begin{equation}
a_j=\sum_{k=0}^{N-1}G(t_k)e^{2\pi if_jt_k}
\end{equation}
where $N$ is the total number of sampling points in the time span
$\Delta T$, and $t_k=k\Delta T/N$ and $f_j=j/\Delta T$, and
$|a_j|=|a_{-j}|$ for a real function $G(t)$.  To indicate the
amplitude of the light variation with the frequency $f_j$, we use the
quantity defined by
\begin{equation}
A(\omega_j)=2|a_j|/N,
\end{equation}
where $\omega_j=2\pi f_j$.


\section{Numerical Results}

The surface $r$-modes propagating in the fluid ocean of a rotating
neutron star may be classified in terms of three quantum numbers
$(m,k,n)$, where $m$ is the azimuthal wave number around the rotation
axis, $k$ is the number of latitudinal nodes, and $n$ is the number of
radial nodes of the eigenfunctions.  For the quantum number $k$, we
count the latitudinal nodes of the eigenfunctions ${\rm Re}[\delta
T(R,\theta)]$, and for the quantum number $n$ we count the radial
nodes of the eigenfunction $iT_{l_1}(r)$.  In this paper, we exclude
the $r$-modes of $l^\prime=|m|$ for light curve calculations, since
their frequency changes during bursts are much larger than is
suggested by observations (e.g., Cumming \& Bildsten 2000, Lee 2004).
Then, the $r$-modes with the quantum number $k=0$, 1, 2, $\cdots$
correspond to the $r$-modes of $l^\prime=|m|+1$, $|m|+2$, $|m|+3$,
$\cdots$, respectively, and the corotating frame frequency $\omega$
decreases in this order for given $m$, $n$, and $\Omega$.  The
$r$-modes with even $k$ are denoted ``even'' modes, and have
eigenfunctions $\delta T(R,\theta)$ that are symmetric about the
equator of the star, while the $r$-modes with odd $k$ are called
``odd'' modes and are antisymmetric about the equator.

For light curve calculations in this paper, we consider only the
fundamental $r$-modes with no radial nodes ($n=0$) of the
eigenfunction $iT_{l_j}$, because they are the most strongly excited
in thermonuclear bursts (Lee 2004).  In addition, since the $r$-modes
associated with high $m$ and $k$ are unlikely to produce light
variations with appreciable amplitudes, we restrict our numerical
analysis to the fundamental modes with $m=1$ or $m=2$ and with $k=0$ or
$k=1$.  Employing these simplifications, and assuming two $r$-modes
with different $m$ may be excited, we can express the surface
temperature for light curve calculation as
\begin{equation}
T(R,\theta,\phi,t)=T_0
+{\rm Re}\left[C_1\delta T^{\alpha_1}(R,\theta)
e^{ i(\phi+\sigma_{\alpha_1} t)}+
C_2e^{i\chi}\delta T^{\alpha_2}(R,\theta)
e^{ i(2\phi+\sigma_{\alpha_2} t)}\right],
\end{equation}
where 
\begin{equation}
\delta T^{\alpha_1}(R,\theta)=\sum_{j\ge1}\delta T_{l_j}^{\alpha_1}(R)
Y_{l_j}^1(\theta,0), \quad
\delta T^{\alpha_2}(R,\theta)=\sum_{j\ge1}\delta T_{l_j}^{\alpha_2}(R)
Y_{l_j}^2(\theta,0),
\end{equation}
and $C_1=C_{\alpha_1}$ and $C_2=C_{\alpha_2}$ are here real constants
for the amplitudes of the modes, and $\chi$ is a real parameter,
giving the phase difference between the two $r$-modes.  If we define
$\alpha=(m,k,n)$, we have $\alpha_1=(1,0,0)$ and $\alpha_2=(2,0,0)$
for the even modes, and $\alpha_1=(1,1,0)$ and $\alpha_2=(2,1,0)$ for
the odd modes.  For the oscillation frequencies, we adopt the
approximation that $\sigma_{\alpha_1}=-\Omega$ and
$\sigma_{\alpha_2}=-2\Omega$.  The parameters we need to calculate the
function $G(t)$ defined by equation (9) are the mass $M$, the radius
$R$, the angular rotation speed $\Omega$ of the star, the oscillation
amplitudes $C_1$ and $C_2$, the phase difference $\chi$ between the
two modes, and the inclination angle $i$.  For the function $G_{E_\infty}(t)$
defined by equation (8), we need in addition the surface temperature
$T_0$ and the photon energy $E_\infty$.

\begin{table*}
\centering
\begin{minipage}{140mm}
\caption{Oscillation frequencies $\bar\omega$ and $\bar\sigma/m$ for
the fundamental ($n=0$) $r$-modes for the mass-accreting envelope
model with $\dot M=0.1\dot M_{\rm Edd}$ for $M=1.4M_\odot$ and
$R=10$km. Here the frequencies $\bar\omega$, $\bar\sigma$, and
$\bar\Omega$ are normalized by $\sqrt{GM/R^3}$.}
\begin{tabular}{@{}cccccc@{}}
\hline & & \multicolumn{2}{c}{$\bar\Omega=0.2$} &
 \multicolumn{2}{c}{$\bar\Omega=0.4$}\\ $k$ & $m$
 &$\bar\omega$&$\bar\sigma/m$ &$\bar\omega$&$\bar\sigma/m$ \\ \hline 0
 & 1 &5.31(-3) & -0.1947& 5.31(-3)& -0.3947 \\ & 2 &1.02(-2) &
 -0.1949& 1.04(-2)& -0.3948 \\ 1 & 1 &3.24(-3) & -0.1968& 3.22(-3)&
 -0.3967 \\ & 2 &6.32(-3) & -0.1968& 6.36(-3)& -0.3968 \\ \hline
\end{tabular}
\end{minipage}
\end{table*}

For the $r$-mode calculation, we use a mass-accreting envelope model
with the rate $\dot M=0.1\dot M_{\rm Edd}$ for a neutron star having
$M=1.4M_\odot$ and $R=10^6$cm, where $\dot M_{\rm Edd}=4\pi
cR/\kappa_e$ with $\kappa_e$ being the electron scattering opacity.
This envelope model is a fully radiative model with no convective
regions in it, and the detail of the envelope calculation is given in
Strohmayer \& Lee (1996) and Lee (2004).  To calculate the modes
propagating in the mass-accreting envelope of a rotating neutron star,
we follow the method described by Lee \& Saio (1987, 1993).  Note that
the formulation employed for the envelope and oscillation mode
calculations is Newtonian, and no general relativistic effects are
included.  In Table 1, the oscillation frequencies $\bar\omega$ and
$\bar\sigma/m$ are tabulated for the fundamental $r$-modes with low
$m$ and $k$ for $\bar\Omega=0.2$ and $\bar\Omega=0.4$, 
where the frequencies $\bar\omega$, $\bar\sigma$, 
and $\bar\Omega$ are normalized frequencies
by $\sqrt{GM/R^3}$. It is
important to note that the corotating frame oscillation frequency
$\bar\omega$ is almost insensitive to $\bar\Omega$ for rapidly
rotating stars (Lee 2004), and that for a given $k$ the inertial frame
pattern speed $\bar\sigma/m$ for $m=2$ is nearly equal to that for
$m=1$, which may justify the simplification of $\sigma=-m\Omega$.  The
temperature variations $\delta T(R,\theta)/T_0$ caused by the surface
$r$-modes for $\bar\Omega=0.2$ are shown in
Figure 1, where panels (a) to (d) are for $\alpha=(1,0,0)$, $(2,0,0)$,
$(1,1,0)$, and $(2,1,0)$, respectively, and the solid and the dashed
lines denote respectively the real and imaginary parts of the function
$\delta T(R,\theta)/T_0$.  Here, the amplitude normalization is given
by $|{\rm Re}[iT_{l^\prime_1}(R)/R]|=1$ at the surface of the star.
We note that the functions $\delta T(R,\theta)/T_0$ and
$iT_{l_1}(R)/R$ have comparable amplitudes to each other at the
stellar surface, and that the imaginary part of $\delta
T(R,\theta)/T_0$ is much smaller than the real part.  The $\theta$
dependence of the function $\delta T(R,\theta)/T_0$ for the even (odd)
$r$-mode with $m=1$ is quite similar to that for the even (odd) $r$
mode with $m=2$ except for the fact that the maxima of the functions
for $m=2$ are larger than those for $m=1$.  In this paper, the
temperature perturbations $\delta T(R,\theta)$ calculated for $R=10$km
will be used for different values of $R$ for simplicity.  Applying the
approximation $\sigma=-m\Omega$, we have $m\phi+\sigma_{\alpha}t=m\phi
-m\Omega t$, and we call the Fourier amplitudes $A(\omega_j)$ at
$\omega_j=\Omega$ and $\omega_j=2\Omega$ the fundamental and the first
harmonic components, respectively.

Since we are applying a linear theory of stellar pulsations, we cannot
self-consistently determine the amplitudes of the modes, which forces
us to treat the oscillation amplitudes as parameters in the modeling.
For simplicity, we use $C_1=0.2$ for $m=1$ and $C_2=0.2$ for $m=2$ for
the mode amplitudes throughout the following calculations.  Since the
toroidal component of the displacement vector is dominant for the
$r$-modes, the velocity, $v$, of a surface fluid element may be
approximately given by $v\sim |\omega iT_{l_1}(R)|=R\sqrt{GM/R^3}
|\bar\omega|\times|iT_{l_1}(R)/R|$.  Since $|\bar\omega|\ltsim 0.01$,
we have the fluid velocity $v\ltsim 10^{-3}c$ for $|iT_{l_1}(R)/R|\sim
0.2$ if we adopt the stellar parameters $M=1.4M_\odot$ and $R=10$km,
for which $R\sqrt{GM/R^3}\sim 0.5c$.  If we use the sound velocity
$v_s= k_BT/\mu H$ at the surface, we have $v_s\sim 10^{-3}c$ for
$T=10^7$K and $\mu=0.5$, where $H$ and $\mu$ denote the hydrogen mass
and the mean molecular weight, respectively.  For the value of
$C_1=0.2$ (or $C_2=0.2$) for the amplitude parameter, the fluid
velocity at the surface is much smaller than the velocity of light and
is comparable to or less than the local sound velocity at the surface.

\subsection{Light curves produced by a single $r$-mode with low $m$}

Since the imaginary part of $\delta T(R,\theta)/T_0$ is much smaller
than the real part, assuming that only a single $r$-mode of low $m$ is
excited, we may simply write the surface temperature as
\begin{equation}
T=T_0\left[1+C(\theta)\cos (m\phi-m\Omega t)\right],
\end{equation}
where $C(\theta)={\rm Re}[C_j \delta T(R,\theta)/T_0]$ and $j=1$ or $j=2$.
Note that setting the origins of times $t$ and $t_\infty$ appropriately,
we can assume $\Omega t=\Omega_\infty t_\infty$.
The function $G(t)$ is given by
\begin{equation}
G(t)=G_0+G_1^c\cos m\Omega t+G_1^s\sin m\Omega t
+G_2^c\cos 2m\Omega t+G_2^s\sin 2m\Omega t+\cdots,
\end{equation}
where
\begin{equation}
G_0=\int_0^12qdq\int_0^{2\pi}{d\phi^\prime\over 2\pi}
{1+1.5C^2(\theta)\over f^3},
\end{equation}
\begin{equation}
G_1^c=\int_0^12qdq\int_0^{2\pi}{d\phi^\prime\over 2\pi}
3f^{-3}C(\theta)\left[1+{C^2(\theta)\over 4}\right]\cos m(\phi+\Delta\phi),
\end{equation}
\begin{equation}
G_1^s=\int_0^12qdq\int_0^{2\pi}{d\phi^\prime\over 2\pi}
3f^{-3}C(\theta)\left[1+{C^2(\theta)\over 4}\right]\sin m(\phi+\Delta\phi),
\end{equation}
\begin{equation}
G_2^c=\int_0^12qdq\int_0^{2\pi}{d\phi^\prime\over 2\pi}
{3\over 2}f^{-3}C^2(\theta)\cos 2m(\phi+\Delta\phi),
\end{equation}
\begin{equation}
G_2^s=\int_0^12qdq\int_0^{2\pi}{d\phi^\prime\over 2\pi}
{3\over 2}f^{-3}C^2(\theta)\sin 2m(\phi+\Delta\phi),
\end{equation}
where $\Delta\phi=\Omega_\infty \Delta t_\infty (b)$, and we have
replaced $t_\infty$ by $t_\infty-\Delta t_\infty(b)$ to calculate the
light curves at $t_\infty$ at the observer.  It may be important to
note that the coefficients $G_1^s$ and $G_2^s$ vanish if we assume
$f=1$ and $\Delta\phi=0$.  Equation (27) indicates that there appears
a periodic component having the frequency $2m\Omega$ unless $G_2^c$
and $G_2^s$ vanish simultaneously.  The terms associated with the
frequency $m\Omega$, for example, can be rewritten as
\begin{equation}
G_1^c\cos m\Omega t+G_1^s\sin m\Omega t=G_1\cos\left(m\Omega t-\delta\right),
\end{equation}
where $G_1=\sqrt{(G_1^c)^2+(G_1^s)^2}$, $\tan\delta=G_1^s/G_1^c$, and
the value of $\Omega t=(\delta+2n\pi)/m$ that maximizes the variation
depends on the parameters $R$ and $i$, where $n$ is an integer.  Since
the imaginary part of $\delta T(T,\theta)$ is negligible compared with
the real part, we have in a good approximation
\begin{equation}
A(m\Omega)=G_1, \quad A(2m\Omega)=G_2,
\end{equation}
where $G_2=\sqrt{(G_2^c)^2+(G_2^s)^2}$, and the Fourier amplitude $A(\omega_j)$
is defined by equation (23).
If we define the phase shift $\delta_j$, using the Fourier transform $a_j$ 
given in equation (22), as
\begin{equation}
\tan\delta_j={\rm Im}(a_j)/{\rm Re}(a_j),
\end{equation}
we have to good approximation $\delta=\delta_j$, selecting the integer
$j$ so that $2\pi f_j=m\Omega$.

The quantities $G_1$ and $\delta/(2m\pi)$ defined by equation (33) are
plotted as a function of the inclination angle $i$ for the even
$(k=0)$ $r$-mode with $m=1$ in Figure 3, and for the even $(k=0)$
$r$-mode with $m=2$ in Figure 4, where we assume $\bar\Omega=0.2$, and
the dotted, short-dashed, solid, long-dashed, and dot-dashed lines
indicate the radius $R=8$, 9, 10, 15, and 20km, respectively.  The
amplitude $G_1$ vanishes at $i=0$ because the temperature variations
are proportional to $e^{im\phi}$ where $\phi=\phi^\prime$ for $i=0$.
For given $i$ and $R$, the amplitude $G_1$ for $m=1$ is usually larger
than that for $m=2$.  For a given inclination angle $i$, the amplitude
$A(m\Omega)=G_1$ decreases as the radius $R$ decreases.  This is
because the maximum angle $\theta^\prime_{\rm max}$ increases with
decreasing $R$ so that almost the whole surface area of the star can
be seen at any time by the distant observer.  For the even modes, the
amplitude $G_1$ for a given $R$ monotonically increases with
increasing $i$.  As shown by Figures 3 and 4 the phase shift
$\delta/(2m\pi)$ for the even $r$-modes are negative.  The absolute
value $|\delta/(2m\pi)|$ for $m=1$, which stays small, increases with
decreasing $R$, and it is rather insensitive to the inclination angle
$i$ for a given $R$.  The absolute value $|\delta/(2m\pi)|$ for $m=2$
also increases as $R$ decreases and becomes as large as
$|\delta/(2m\pi)|\sim 0.2$ for the smallest $R$ considered here.  It
shows a weak dependence on $i$, particularly for the radii $R\sim
10$km.

The quantities $G_1$ and $\delta/(2m\pi)$ are given as a function of
the inclination angle $i$ for the odd $(k=1)$ $r$-mode with $m=1$ in
Figure 5, and for the odd $(k=1)$ $r$-mode with $m=2$ in Figure 6,
where $\bar\Omega=0.2$, and the different curves have the same meaning
as in Figure 3.  For the odd modes, the amplitude $G_1$ vanishes at
$i=90^\circ$ also, because the temperature variation pattern is
antisymmetric about the equator of the star.  As in the case of the
even $r$-modes, the amplitude $G_1$ for the odd $r$-modes decreases
with decreasing $R$ for a given $i$.  Although $\delta/(2m\pi)$ for
larger radii $R\gtsim 15$km stays constant with varing $i$, it shows a
strong dependence on $i$ for smaller radii $R\ltsim 10$km.

Figure 7 shows the amplitude ratio $G_2/G_1$ for the even $r$-mode
with $m=1$ (upper panel) and for the odd $r$-mode with $m=1$ (lower
panel) as a function of $i$.  We find the ratio is at most of order
$\sim$0.01 for the even $m=1$ $r$-modes, which indicates that the
first harmonic component cannot be significant in the light curves
produced by a single even $r$-mode of low $m$.  For the odd $m=1$ $r$
mode, on the other hand, the first harmonic component has appreciable
amplitude compared to the fundamental, particularly when $i\sim
90^\circ$.  Actually, the ratio $G_2/G_1$ diverges as $i$ increases to
$90^\circ$.

As examples, we show as a function of $\Omega t/2\pi$ the light curves
produced by the even, $m=1$ $r$-mode in Figure 8, and by the odd,
$m=1$ $r$-mode in Figure 9, where we assume $\bar\Omega=0.2$ and
$i=60^\circ$.  Figure 10 shows the light curves generated by the odd,
$m=1$ $r$-mode as a function of $\Omega t/2\pi$ for $\bar\Omega=0.2$
and $R=10$km, where the dotted, short-dashed, solid, long-dashed, and
dot-dashed lines indicate the inclination angle $i=10^\circ$,
$30^\circ$, $50^\circ$, $70^\circ$, and $90^\circ$, respectively.  At
$i=90^\circ$, only the first harmonic component associated with
$2m\Omega$ appears.

\subsection{Dependence on $\bar\Omega$}

As $\Omega$ increases, the amplitudes of the $r$-modes are more
strongly confined to the equatorial region of the star, which makes
the amplitude $G_1$ larger for the even $r$-modes but smaller for the
odd modes.  Note that the frequency $\omega$ is only weakly dependent
on $\Omega$ for the modes.  To see the effects of rapid rotation, we
calculate $G_1$ and $\delta$ for the even, $m=1$ $r$-mode for
$\bar\Omega=0.4$, and we plot the results in Figure 11, where the
different curves represent different radii as in Figure 3.  We note
that the magnitude of the phase shift $\delta$ is larger for
$\bar\Omega=0.4$ than for $\bar\Omega=0.2$ although $\delta$ is still
only weakly dependent on the inclination angle $i$ for the mode.

\subsection{Light curves produced by two $r$-modes}

Although the even, $m=1$ $r$-mode will produce the largest light
variations of the modes considered here, the mode is not likely to
produce a first harmonic component $A(2\Omega)$ of appreciable
strength. In order to account for observed light curves that contain a
substantial first harmonic component, we may need to assume that the
$r$-modes with $m=1$ and $m=2$ are excited simultaneously.  As an
example, we calculate the light curves $G(t)$ produced by the two
simultaneously excited even $r$-modes with $m=1$ and $m=2$, assuming
$C_1=C_2=0.2$, $\chi=0$, $i=90^\circ$, and $\bar\Omega=0.2$. We plot
$G(t)$ as a function of $\Omega t/2\pi$ in Figure 12 and the Fourier
amplitudes $A(\Omega)$ and $A(2\Omega)$ as a function of the radius
$R$ in Figure 13, where the different curves in Figure 12 have the
same meaning as in Figure 3.  Obviously, using two $r$-modes with
different $m$'s, we can produce a variety of light curves, including
those with a harmonic content similar to that seen from XTE J1814-338
(Strohmayer et al. 2003). It is important to note that, even if we
abandon the simplification given by $\sigma_{\alpha_1}=-\Omega$ and
$\sigma_{\alpha_2}=-2\Omega$, we have in a good approximation
$2\sigma_{\alpha_1}=\sigma_{\alpha_2}$ as shown by Table 1.

\subsection{Phase Lags}

The amplitudes of the light curves produced by the surface $r$-modes
are inevitably related to the parameters $C_i$'s as well as to the
ratio $R/M$.  Since the oscillation amplitudes are difficult to
determine within the framework of a linear theory of stellar
pulsations and the ratio $R/M$ itself is one of our main parameters to
determine observationally, it would be useful to find an additional
observable that can be used as an indicator for the ratio $R/M$.
Considering the even $r$-mode with $m=1$, which is most likely to be
responsible for the light variations, we note that the phase shift
$\delta$, which is almost independent of the oscillation amplitude
parameter $C_1$, is rather insensitive to the inclination angle $i$
but is almost a monotonic function of the radius $R$ (i.e., the ratio
$R/M$) for a given $i$.  If it is true that the observed light
variations during burst oscillations are caused by an even, $m=1$
$r$-mode, it would be possible to use the phase shift $\delta$ as an
indicator for the ratio $R/M$.  To derive an observable using the
phase shift $\delta$, we calculate the function $G_{E_\infty}(t)$
defined in equation (8) for appropriate values of $T_0$ and
$E_\infty$, and Fourier analyze the function $G_{E_\infty}(t)$ to
obtain the phase shift $\delta_j$ at $\omega_j=\Omega$ using equation
(35), which we denote $\delta_E$.  In Figure 14, we plot as a function
of the inclination angle $i$ the difference $\delta-\delta_{E_0}$ for
the even, $m=1$ $r$-mode (upper panel) and for the odd, $m=1$ $r$-mode
(lower panel) where we have assumed $\bar\Omega=0.2$,
$E_\infty=E_0=1$kev, $k_BT_0=2.3$kev, and the various curves represent
different radii as in Figure 3.  Since the difference
$\delta-\delta_{E_0}$ for the even $r$-mode with $m=1$ is almost
insensitive to $i$ and its magnitude increases almost monotonically
with the radius $R$ for a given $i$, the difference may be useful as
an observational indicator for the ratio $R/M$.  On the other hand,
the difference for the odd mode shows a strong dependence on $i$ for
the radius $R\ltsim 10$km.  Since the phase shifts $\delta$ found in
this paper are due mainly to the Doppler effects associated with rapid
rotation of the star, the magnitude of the difference will be large
for higher $\bar\Omega$ for given $M$ and $R$.

As shown in Figure 14, the phase difference, $\delta-\delta_{E_0}$,
for the even mode is always negative, but that for the odd mode can be
positive for radii $R\ltsim 10$km. Here, negative and positive values
of $\delta-\delta_{E_0}$ indicate hard leads and lags, respectively,
for $E_\infty=E_0=1$kev and $k_BT_0=2.3$kev.  To compare with
observations more directly, it is convenient to give the phase shift
difference, $\delta_E-\delta_{E_0}$, between the phase for
$E_\infty=E$ and that for a particular $E_\infty=E_0$.  Examples are
given as a function of the photon energy $E$ in Figure 15 for the even
$r$-mode with $m=1$ for $i=60^\circ$ and in Figure 16 for the odd
$r$-mode with $m=1$ for $i=30^\circ$ (upper panel) and for
$i=60^\circ$ (lower panel), where we have used $E_0=1$kev and
$\Omega_\infty/2\pi=400$Hz.  For the even mode we always have negative
$\delta_E-\delta_{E_0}$, indicating hard leads (soft lags), as
suggested by Figure 14.  On the other hand, for the odd $m=1$
$r$-mode, we have both hard leads (soft lags) and hard lags, the
latter of which occur at smaller radii (i.e., smaller $R/M$).
Observationally, there is an indication for hard lags during some
burst oscillations, however, others are consistent with no phase
variations with energy (Muno et al. 2003). Since the even, $m=1$
$r$-mode generates hard leads (as the hot spot model does), our
results suggest that a single even $r$-mode is probably inconsistent
with burst oscillations which show hard lags. However, it might be
possible for an oscillation produced by contributions from both even
and odd modes to match the observed phase lag properties of burst
oscillations.  We will explore detailed modeling and fitting,
including the effects of the detection process, in a subsequent paper.

\section{Discussion}

The surface $r$-modes considered in this paper belong to a subclass of
the equatorial waves (Pedlosky 1987).  In the equatorial $\beta$ plane
approximation, we employ a Cartesian coordinate system in which the
coordinate origin is at the equator and the $x$, $y$, and $z$ axes are
in the eastward, northward, and upward directions, respectively.  The
wave equations in the $\beta$ plane approximation are solved by
applying the method of separation of variables, in which a separation
constant $\lambda$ is introduced between $(x,y,t)$ and $z$.  The
separation constant $\lambda$ is an eigenvalue that is determined by
solving the wave equation in the vertical direction with appropriate
boundary conditions, and reflects the vertical structure of the thin
fluid envelope.  Assuming the perturbed velocity $v^\prime_y(x,y,z,t)$
is given by $v^\prime_y\propto e^{i(kx+\omega t)}\Psi(y)V(z)$, we
obtain for the equatorial waves the dispersion relation given by
(Pedlosky 1987)
\begin{equation}
\lambda\tilde\omega^2+\tilde k/\tilde\omega-\tilde k^2=(2j+1)\sqrt{\lambda},
\end{equation}
where 
\begin{equation}
\tilde\omega=\omega/(\beta_0 L_e), \quad \tilde k=L_e k, \quad 
L_e=\sqrt{N_0D/\beta_0}, \quad \beta_0=2\Omega/R,
\end{equation}
and $N_0$ is a characteristic value 
of the Brunt-V\"ais\"al\"a frequency,
$D$ is the depth of the fluid ocean, and the function $\Psi(y)$ is given by
\begin{equation}
\Psi(y)=\psi_j(\tau)\equiv e^{-\tau^2/2}H_j(\tau)/ \sqrt{2^jj!\pi^{1/2}},
\end{equation}
where $H_j(\tau)$ is the Hermite polynomial,
$\tau=\lambda^{1/4}y/L_e$, and $j$ is an integer.  Here, the function
$V(z)$ is determined by the wave equation in the vertical direction.
The pressure perturbation for the equatorial waves for $j\ge 1$ is
then given by (Pedlosky 1987)
\begin{equation}
p^\prime(x,y,z,t)=- {\rm Re}\left\{ {iA_j\over \lambda^{3/4}}
\left[-\left({j\over 2}\right)^{1/2}
{\psi_{j-1}(\tau)\over\tilde\omega-\tilde k/\sqrt{\lambda}}
+\left({j+1\over 2}\right)^{1/2}
{\psi_{j+1}(\tau)\over\tilde\omega+\tilde k/\sqrt{\lambda}} 
\right]e^{i(kx+\omega t)}V(z)\right\},
\end{equation}
where $A_j$ is an arbitrary constant.
Since the low frequency equatorial waves at rapid rotation have (Lee 2004)
\begin{equation}
\omega\sim{mN_0D/R\over (2j+1)\sqrt{\lambda}},
\end{equation}
where we have assumed $k\sim m/R$, $p^\prime(x,y,z,t)$ reduces to for $j=1$
\begin{equation}
p^\prime(x,y,z,t) \sim - {\rm Re}\left\{ {3\over
2\sqrt{2}\pi^{1/4}}{iA_1\over \lambda^{1/4}L_e k} \left(\tau^2+{1\over
2}\right)e^{-\tau^2/2}e^{i(kx+\omega t)}V(z)\right\},
\end{equation}
and for $j=2$ 
\begin{equation}
p^\prime(x,y,z,t) \sim - {\rm Re}\left\{ {5\over
3\sqrt{2}\pi^{1/4}}{iA_2\over \lambda^{1/4}L_e k}
\tau^3e^{-\tau^2/2}e^{i(kx+\omega t)}V(z)\right\}.
\end{equation}
Knowing that the perturbed potential temperature $\Theta^\prime$ is
approximately given by $\Theta^\prime=\partial p^\prime/\partial z$
(Pedlosky 1987), we find that the latitudinal ($y$) dependence of
$\Theta^\prime(x,y,z,t)$ well reproduces that of $\delta
T(R,\theta)/T_0$ in Figure 2 both for the even $r$-modes ($j=1$) and
for the odd $r$-modes ($j=2$).  Note that the solutions having the
frequency (40) correspond to the type 2 solutions discussed by
Longuet-Higgins (1968).

The oscillation energy $E_W$ of a mode observed in the corotating frame
may be given by (e.g., Unno et al. 1989)
\begin{equation}
E_W={\omega^2\over 2}\int_{\Delta M_e}\pmb{\xi}\cdot\pmb{\xi}^*dM_r
=\bar\omega^2{GM\Delta M_{o}\over R},
\end{equation}
where 
\begin{equation}
\Delta M_o={1\over 2}
\int_{\Delta M_e}{\pmb{\xi}\cdot\pmb{\xi}^*\over R^2}dM_r,
\end{equation}
and $\pmb{\xi}^*$ is the complex conjugate of $\pmb{\xi}$, 
and $\Delta M_e$ is the envelope mass.
If we let $\Delta E_b$ and $\Delta E_p$
denote the energies released during a burst and a persistent phase, 
respectively, we may assume
$\alpha\equiv \Delta E_p/\Delta E_b\sim 10^2$ 
and $\Delta E_p\sim \epsilon GM\Delta M_a/R$, where $\epsilon\sim 0.1$ is a factor
representing the efficiency of energy transformation from gravitational to 
radiation energies and $\Delta M_a=\dot M\Delta t_p$ 
with $\dot M\sim 10^{16}{\rm gs}^{-1}$ being a typical mass accretion rate 
in close binary systems
is the amount of mass accreting during 
the persistent phase $\Delta t_p\sim 10^4$s
(see, e.g., Frank, King, \& Raine 2002).
We then have
\begin{equation}
E_W=\alpha\epsilon^{-1}\bar\omega^2(\Delta M_o/\Delta M_a)\Delta E_b,
\end{equation}
which may be rewritten with appropriate normalization as
\begin{equation}
E_W\sim 2\times {\alpha\over 100}\left({\epsilon\over 0.1}\right)^{-1}
\left({\dot M\over 5\times 10^{-18}M_\odot {\rm s}^{-1}}\right)^{-1}
\left({\Delta t_p\over 10^4 {\rm s}}\right)^{-1}
\left({\bar\omega\over 0.01}\right)^2\left({C\over 0.1}\right)^2
{\left<\Delta  M_o\right>\over 10^{-10}M_\odot}\Delta E_b,
\end{equation}
where we have defined 
\begin{equation}
\left<\Delta M_o\right>={1\over 2}
\int_{\Delta M_e}{\bar{\pmb{\xi}}\cdot\bar{\pmb{\xi}}^*\over R^2}dM_r,
\end{equation}
assuming $\pmb{\xi}=C\bar{\pmb{\xi}}$ with $\bar{\pmb{\xi}}$ being the  
displacement vector normalized as $|{\rm Re}[iT_{l^\prime_1}(R)/R]|=1$.
For the $r$-modes discussed in this paper, we have the ratio 
$\left<\Delta M_o\right>/10^{-10}M_\odot\sim 0.1$ for the even $r$-modes, and
$\left<\Delta M_o\right>/10^{-10}M_\odot\sim 0.01$ for the odd $r$-modes,
for $\bar\Omega=0.2$ and $\Delta M_e=10^{-10}M_\odot$.
Although the estimation given above contains many parameters, all of which are
not necessarily well determined, we may expect that 
the amount of energy required to excite the modes to appreciable amplitudes is 
a modest fraction of that released in a burst.
It is interesting to note that, although the odd $r$-modes need to have
larger amplitudes to produce appreciable light variations 
than the even $r$-modes, the former are easier to excite than the latter for
a given amount of energy released in a burst.

\section{Conclusions}

We have calculated light curves produced by the surface $r$-modes of a
rotating neutron star, taking account of the effects of gravitational
light bending, gravitational red shift, and the difference in the
arrival times of photons traveling in the static Schwarzschild
spacetime, as well as the Doppler shift of photon energy due to rapid
spin of the star.  We find that the fundamental even $r$-mode with
$m=1$ produces the largest light variations.  The light curves
produced by a single even $r$-mode of a given $m$ are dominated by the
fundamental component with frequency $m\Omega$, although those
produced by a single odd $r$-mode also contain a first harmonic
component of appreciable amplitude as well as the fundamental
component, the relative strengths depending on the inclination angle
$i$.  The phase shift $\delta-\delta_{E_0}$ produced by the even $m=1$
$r$-mode is only weakly dependent on the inclination angle $i$ and is
almost a monotonic function of $R$ for a given $i$.  The phase shift
produced by the even $r$-mode with $m=1$ is a hard lead.  The phase
shift produced by the odd, $m=1$ $r$-mode, on the other hand, depends
both on $i$ and on $R$, and there exists a parameter space of $R$ and
$i$ which produces hard lags.

It is useful to translate the phase shift difference into a hard-lag
measure, $\Delta t_{\rm Lag}$, which is given for
$\Omega_\infty/2\pi=400$Hz by
\begin{equation}
\Delta t_{\rm Lag}={2\pi\over\Omega_\infty}{\delta_E-\delta_{E_0}\over 2m\pi}=
2.5\times{\delta_E-\delta_{E_0}\over 2m\pi}\quad {\rm ms},
\end{equation}
where negative $\Delta t_{\rm Lag}$ means hard leads.  As indicated by
Figure 15, to fit the observed hard leads $\sim -200\mu$s at
$E_\infty\sim 10$kev for SAX J1808.4-3658 (Cui et al. 1998) in terms of
the $even$ $m=1$ $r$-mode we need a little bit smaller values of the
ratio $R/M$.  In fact, Weinberg et al. (2001) have assumed
$M=2.2M_\odot$ and $R=10$km for their fit, which is equivalent to
$R/R_G=1.54$, the value of which is substantially smaller than the
smallest value $R/R_G=1.94$ for $M=1.4M_\odot$ and $R=8$km we use in
this paper.  As suggested by Figure 16, however, the phase shift
difference produced by the $odd$ $m=1$ $r$-mode may give a consistent
value $\sim -200\mu$s at $E_\infty\sim 10$kev for $M=1.4M_\odot$,
$R=10$km, and $i=60^\circ$, although the strong saturation of $\Delta
t_{\rm Lag}$ above $E\gtsim 10$kev is not necessarily well reproduced.

Compared with the hot spot model, the wave model generally produces
smaller amplitude light variations. This is mainly because the entire
surface of the neutron star produces X-ray emission.  On the other
hand, purely sinusoidal light curves observed for most of the burst
oscillators (Muno et al. 2002b) can be explained by assuming the
dominance of a single $r$-mode of a given $m$, and the existence of an
appreciable first harmonic component may be attributable to the
coexistence of a mode with $2m$, or perhaps a superposition of even
and odd modes of different amplitude.  As in the case of the hot spot
model, the wave model also gives hard leads in most of the parameter
space, which may be in conflict with the suggested hard lags seen in
some of the burst oscillators, but might be applicable to the hard
leads observed from the millisecond X-ray pulsars SAX J1808.4-3658,
and IGR J00921+5934 (e.g., Cui et al. 1998; Galloway et al. 2005). The
parameters employed in the light curve calculation include the mass
$M$, the radius $R$, and the spin frequency $\Omega$ of the star, the
inclination angle $i$, the mode indices $(m,k,n)$, and the oscillation
amplitudes $C_j$.  Even if we assume that only the fundamental $n=0$
$r$-modes with $k=0$ or $k=1$ play a role in producing the observed
brightness variations and that the frequency that appears in the light
curves is the spin frequency of the star, we still have several free
parameters that affect the amplitude determination of the light
curves.  Perhaps the most crucial parameters are the oscillation
amplitudes, $C_j$, themselves, which are difficult to determine within
the framework of linear pulsation theory.  Further theoretical studies
combined with model fitting to the observed properties of burst
oscillations will be definitely desired for a better understanding of
the burst oscillations and the underlying neutron stars.

\newpage


\begin{figure}
\centering
\epsfig{file=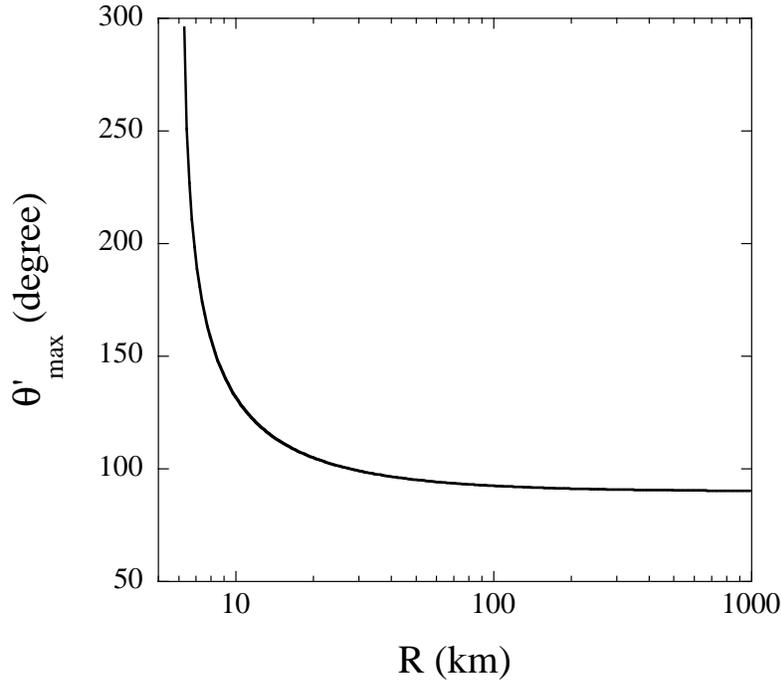,width=0.6\textwidth}
\caption{ $\theta^\prime_{\rm max}$ as a function of the radius $R$ 
for the star with the mass $M=1.4M_\odot$}
\end{figure}

\begin{figure}
\centering
\epsfig{file=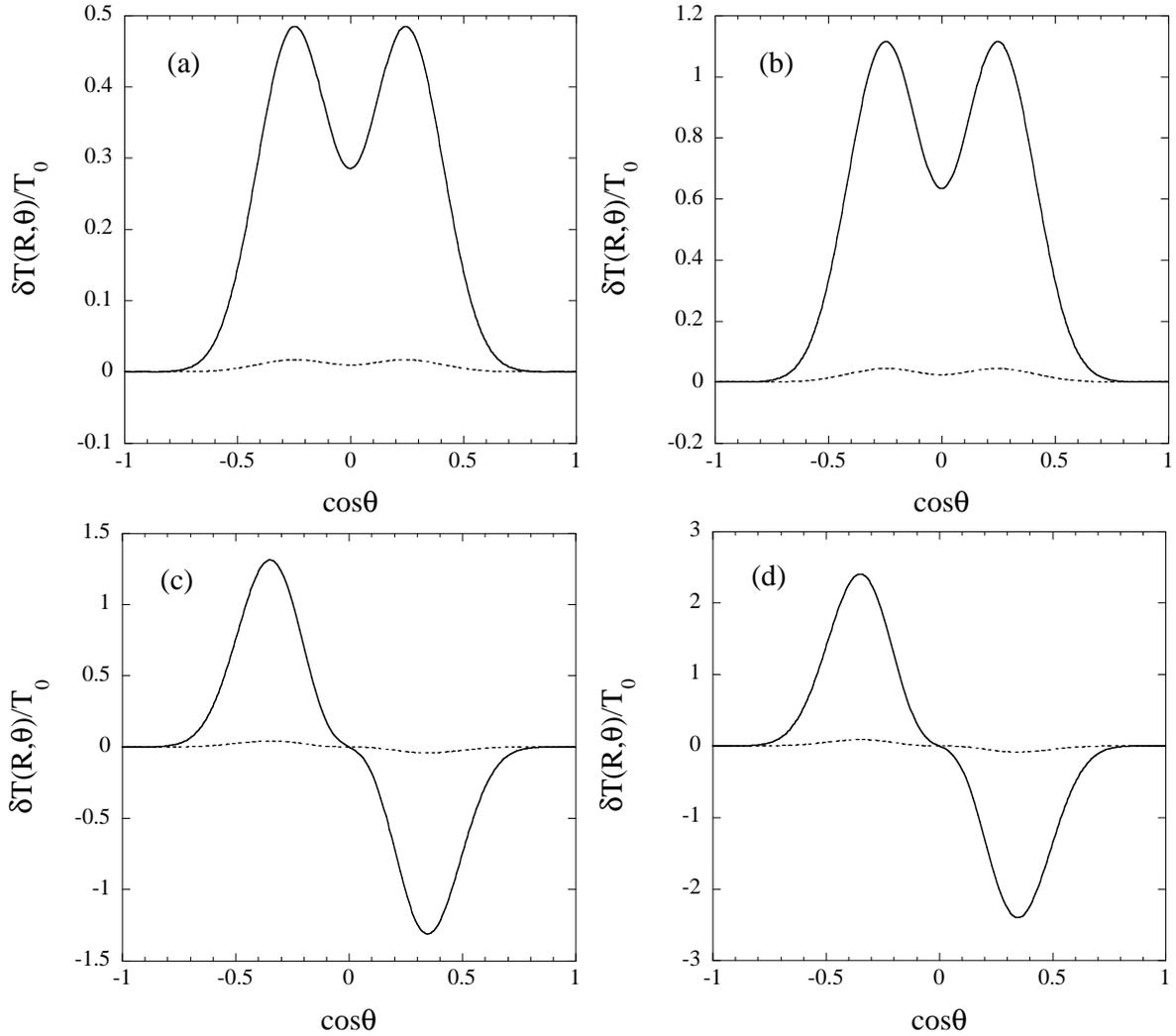,width=0.9\textwidth}
\caption{$\delta T(R,\theta)/T_0$ as a function of $\cos\theta$ for
$\bar\Omega=0.2$.  Panel (a) for the fundamental even $r$-mode with
$m=1$, (b) for the fundamental even $r$-mode with $m=2$, (c) for the
fundamental odd $r$-mode with $m=1$, (d) for the fundamental odd $r$
mode with $m=2$, where the solid and the dotted lines indicate the
real and the imaginary parts, respectively. Here, the amplitude
normalization for the mode is given by $|{\rm Re}[iT_{l_1}(R)/R]|=1$
at the surface.}
\end{figure}

\begin{figure}
\centering
\epsfig{file=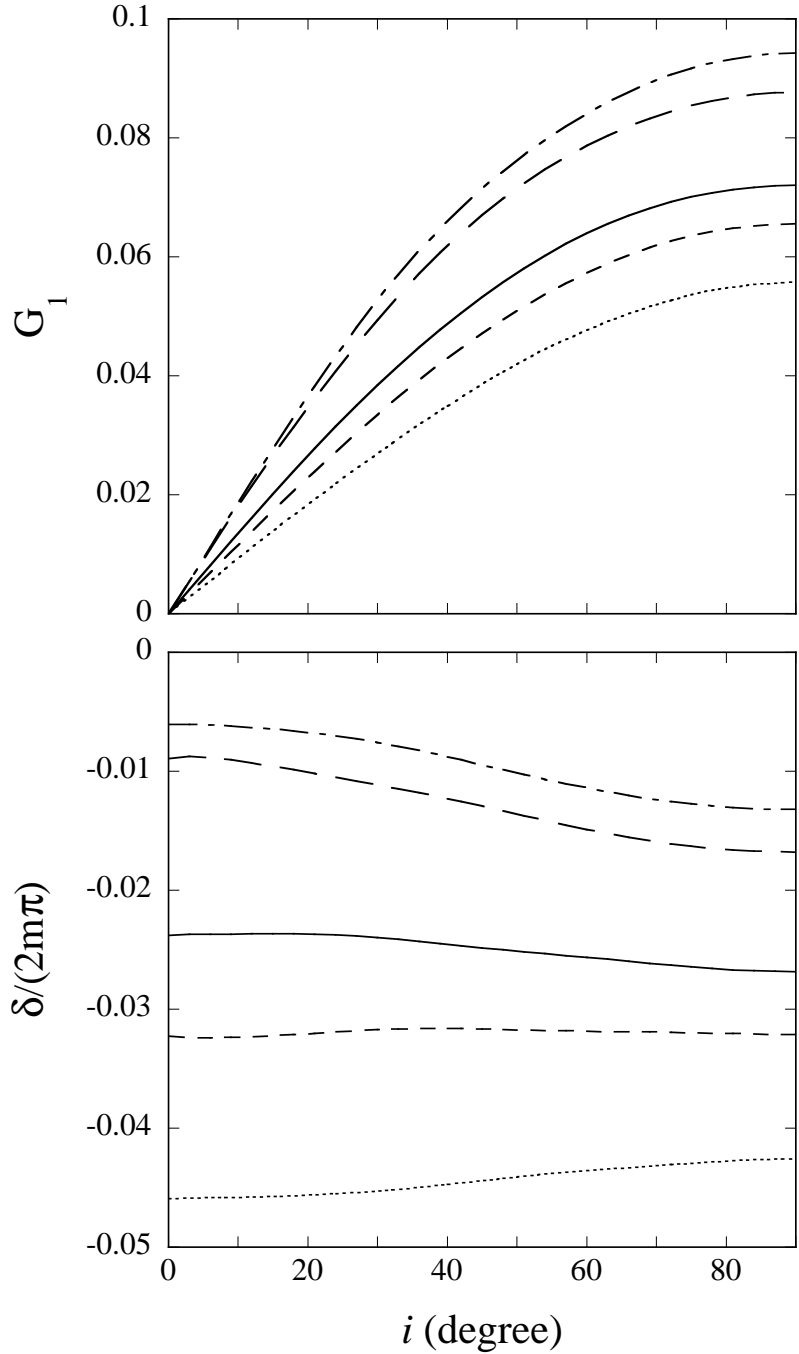,width=0.6\textwidth}
\caption{Amplitude $G_1$ and phase shift $\delta$ as a function of the
inclination angle $i$ for the fundamental even $r$-mode with $m=1$
($C_1=0.2$, $C_2=0$, and $\chi=0$), 
where $\bar\Omega=0.2$, and the dotted, short-dashed,
solid, long-dashed, and dash-dotted lines are for the radii $R=8$, 9,
10, 15, and 20km, respectively.  Here, the amplitude normalization for
the mode is given by $|{\rm Re}[iT_{l_1}(R)/R]|=1$ at the surface.}
\end{figure}

\begin{figure}
\centering
\epsfig{file=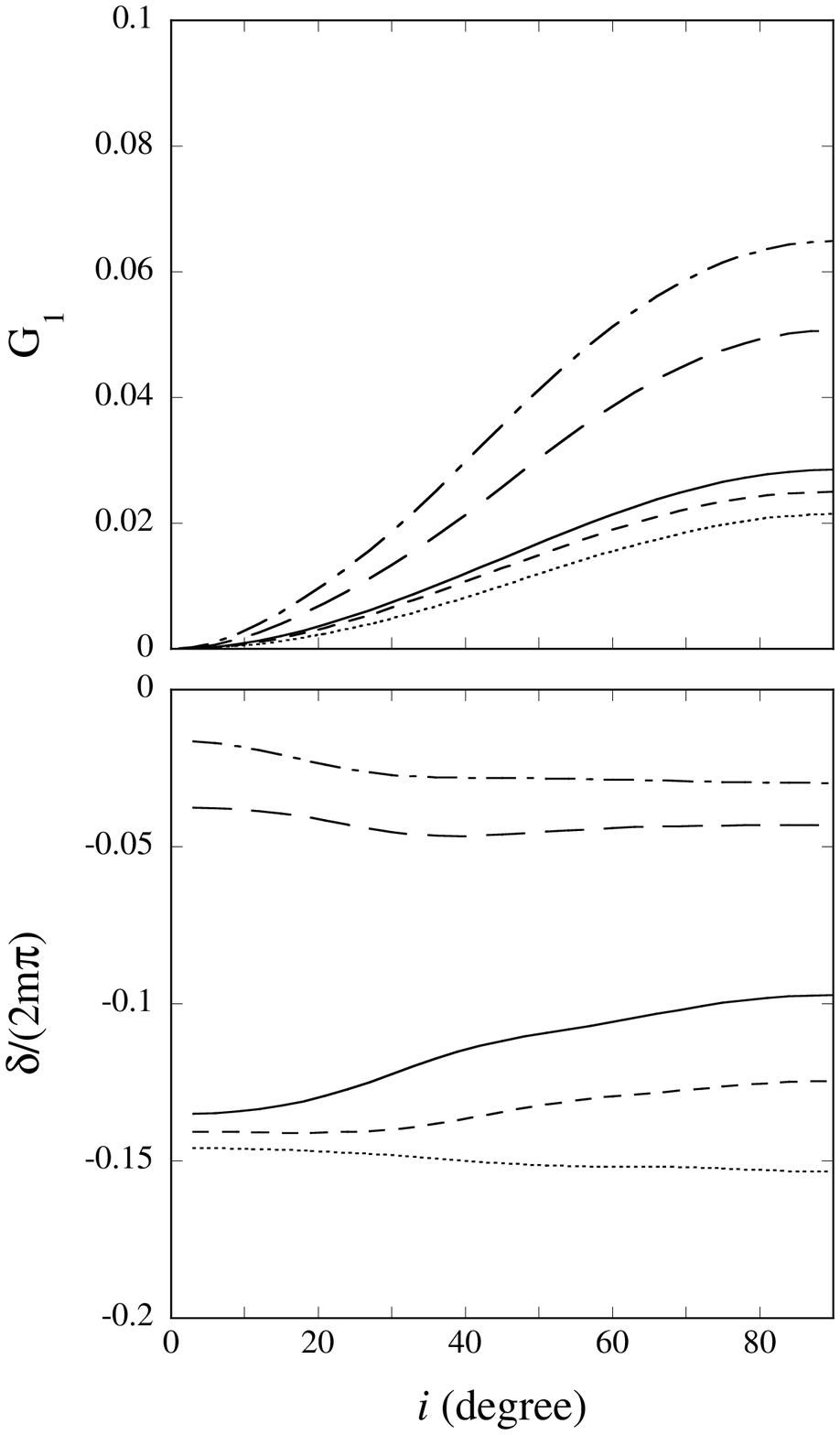,width=0.6\textwidth}
\caption{Same as Figure 3 but for the fundamental even $r$-mode with
$m=2$ ($C_1=0$, $C_2=0.2$, and $\chi=0$).}
\end{figure}

\begin{figure}
\centering \epsfig{file=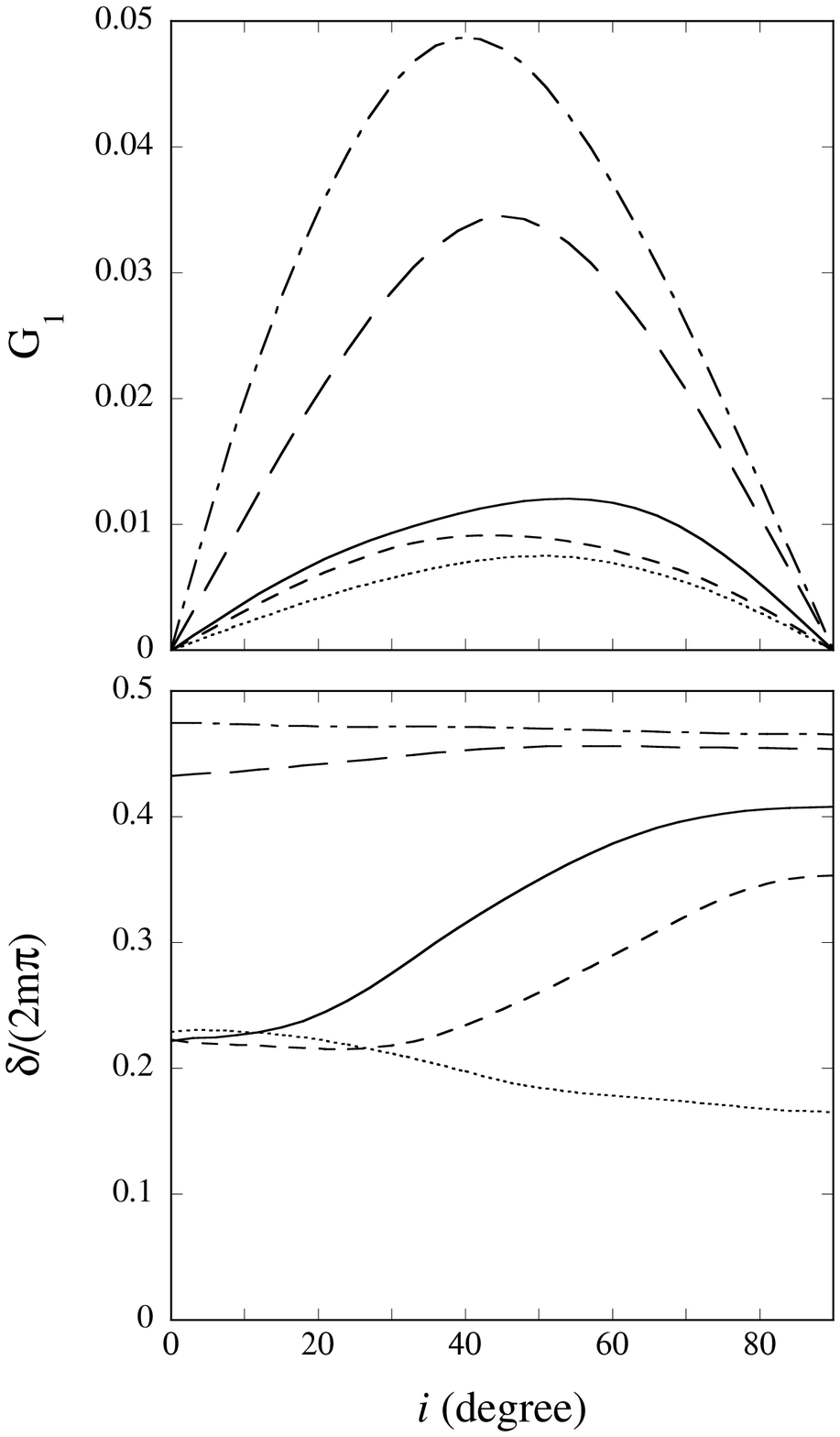,width=0.6\textwidth}
\caption{Same as Figure 3 but for the fundamental odd $r$-mode with
$m=1$ ($C_1=0.2$, $C_2=0$, and $\chi=0$).}
\end{figure}

\begin{figure}
\centering
\epsfig{file=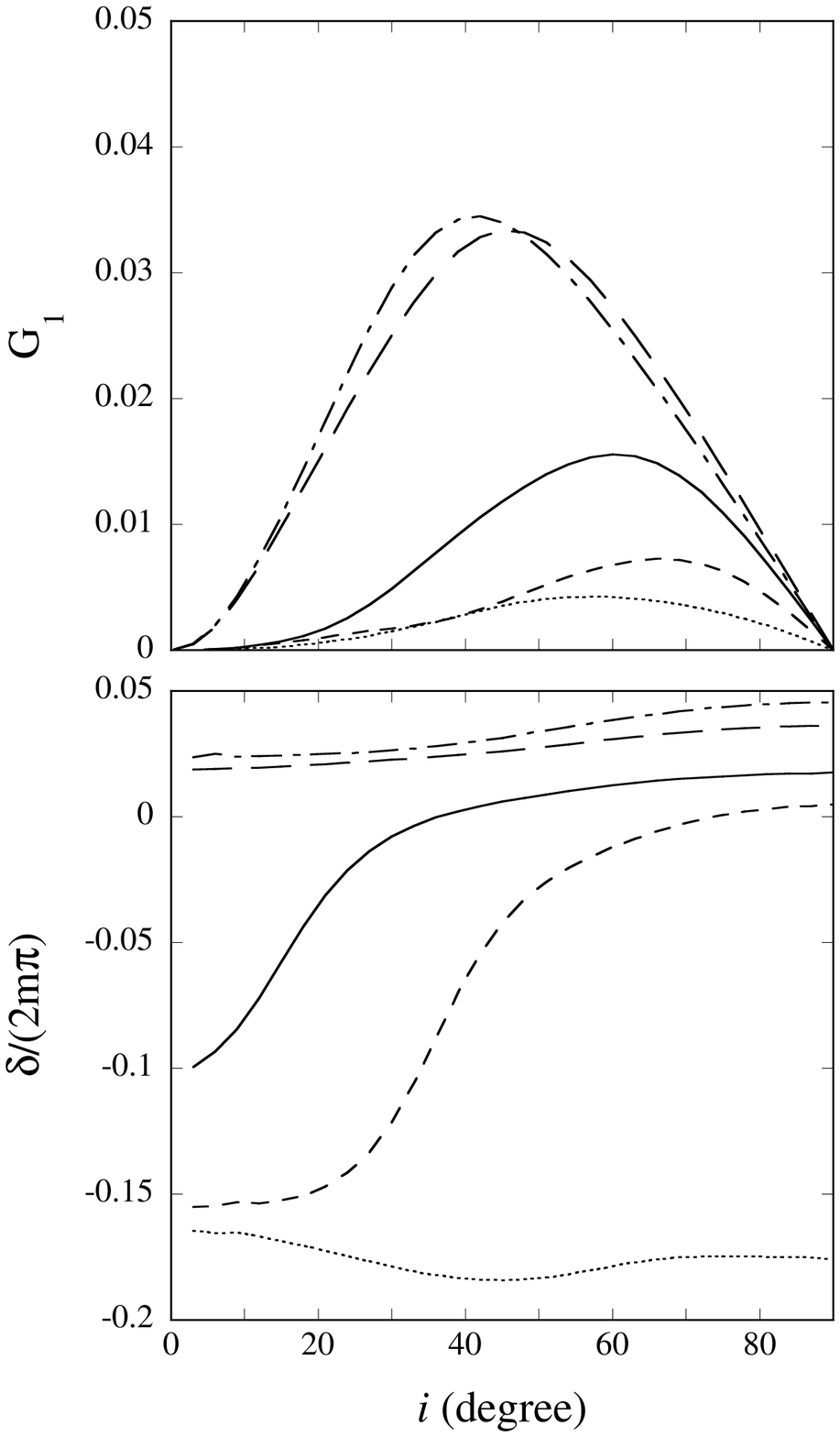,width=0.6\textwidth}
\caption{Same as Figure 3 but for the fundamental odd $r$-mode with
$m=2$ ($C_1=0$, $C_2=0.2$, and $\chi=0$).}
\end{figure}

\begin{figure}
\centering
\epsfig{file=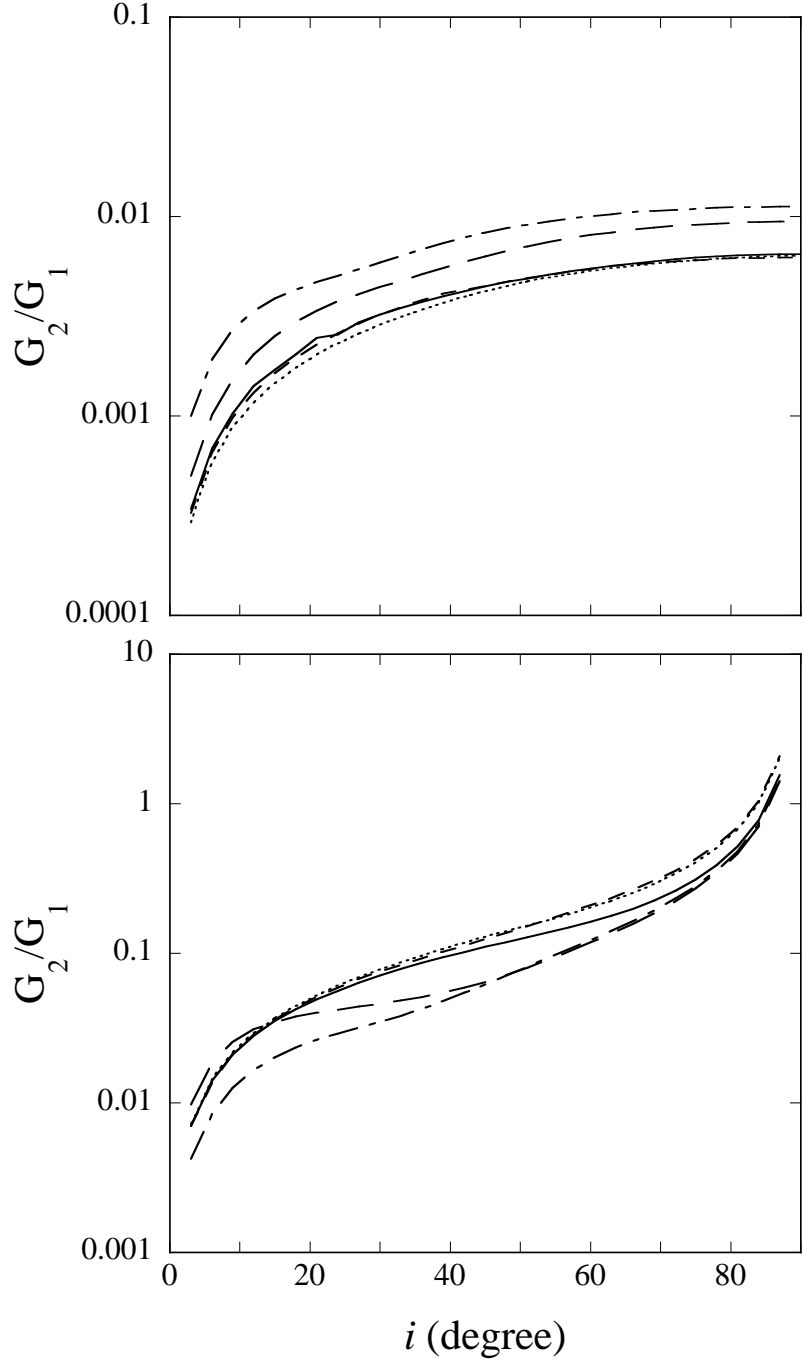,width=0.6\textwidth}
\caption{Ratio $G_2/G_1$ as a function of the inclination angle $i$
for the fundamental even $r$-mode with $m=1$ (upper panel) and for the
fundamental odd $r$-mode with $m=1$ (lower panel), where
$\bar\Omega=0.2$, and the various curves represent different radii as in Figure 3.  
Here, the amplitude normalization for the mode is given
by $|{\rm Re}[iT_{l_1}(R)/R]|=1$ at the surface.}
\end{figure}

\begin{figure}
\centering
\epsfig{file=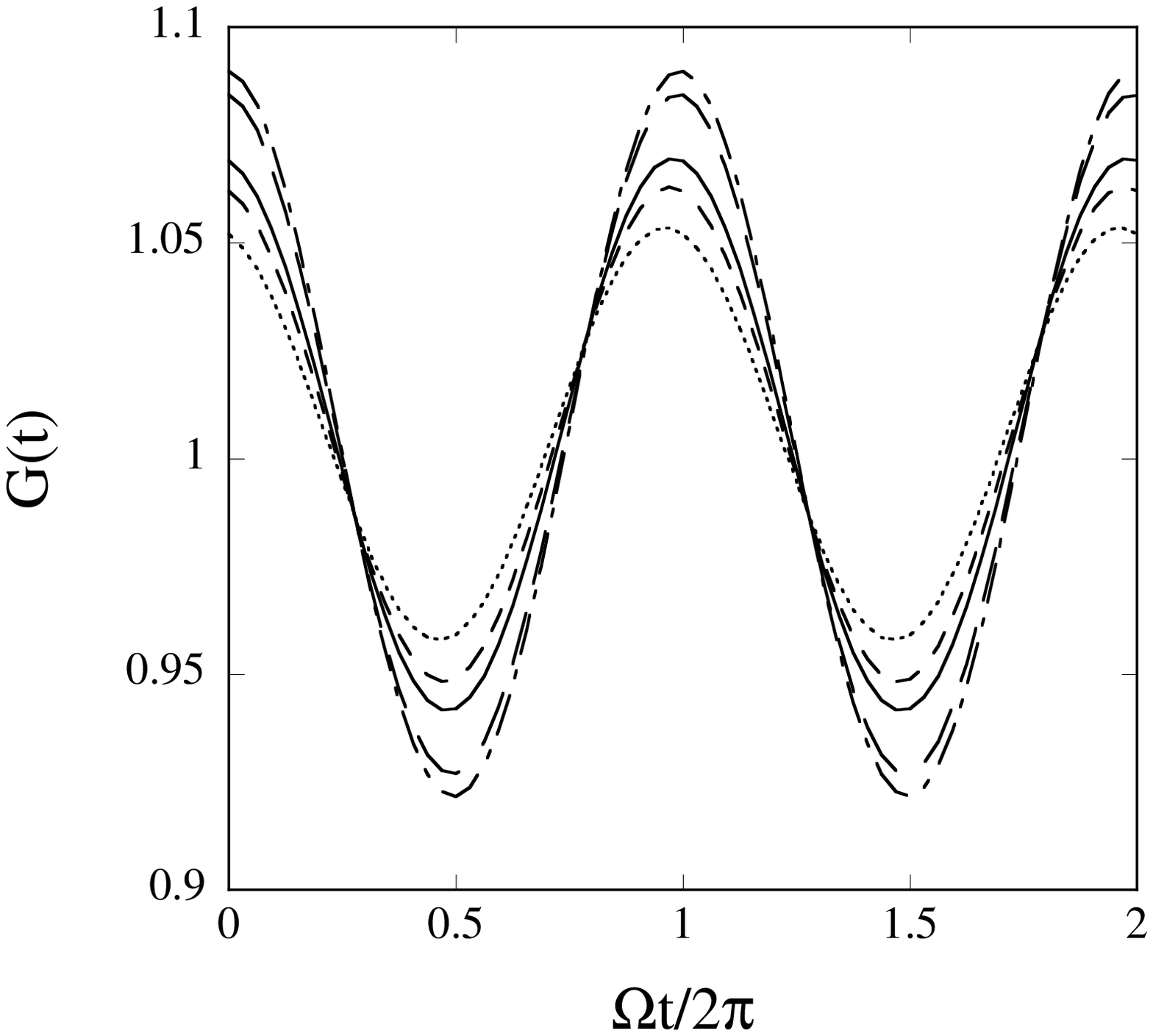,width=0.6\textwidth}
\caption{$G(t)$ as a function of $\Omega t/2\pi$ for the fundamental
$even$ $r$-mode with $m=1$ for $i=60^\circ$, where $C_1=0.2$, $C_2=0$, 
$\chi=0$, and
$\bar\Omega=0.2$, and the various curves represent different radii as in Figure 3.  
Here, the amplitude normalization for the mode is given
by $|{\rm Re}[iT_{l_1}(R)/R]|=1$ at the surface.}
\end{figure}

\begin{figure}
\centering
\epsfig{file=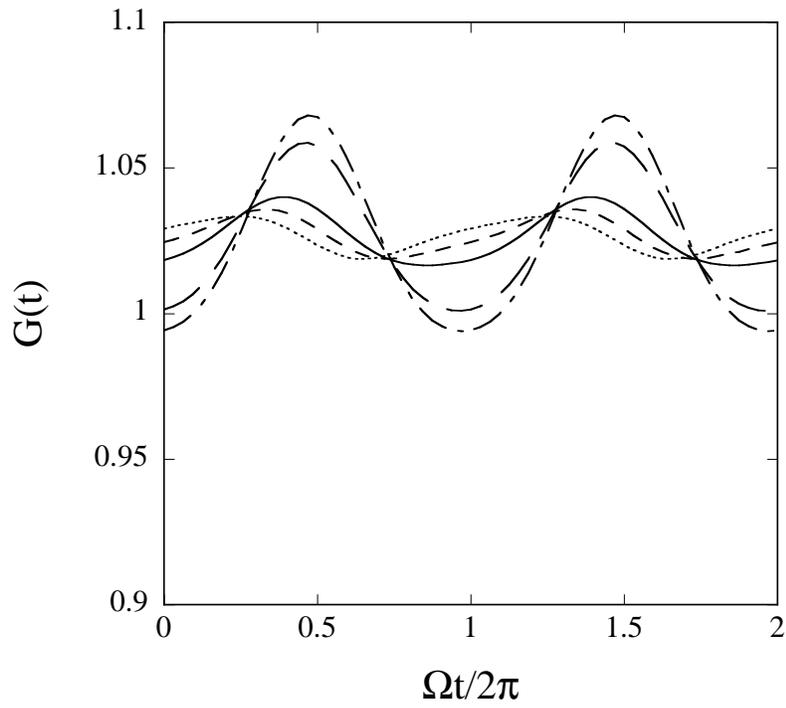,width=0.6\textwidth}
\caption{Same as Figure 8 but for the fundamental $odd$ $r$-mode with $m=1$.}
\end{figure}

\begin{figure}
\centering
\epsfig{file=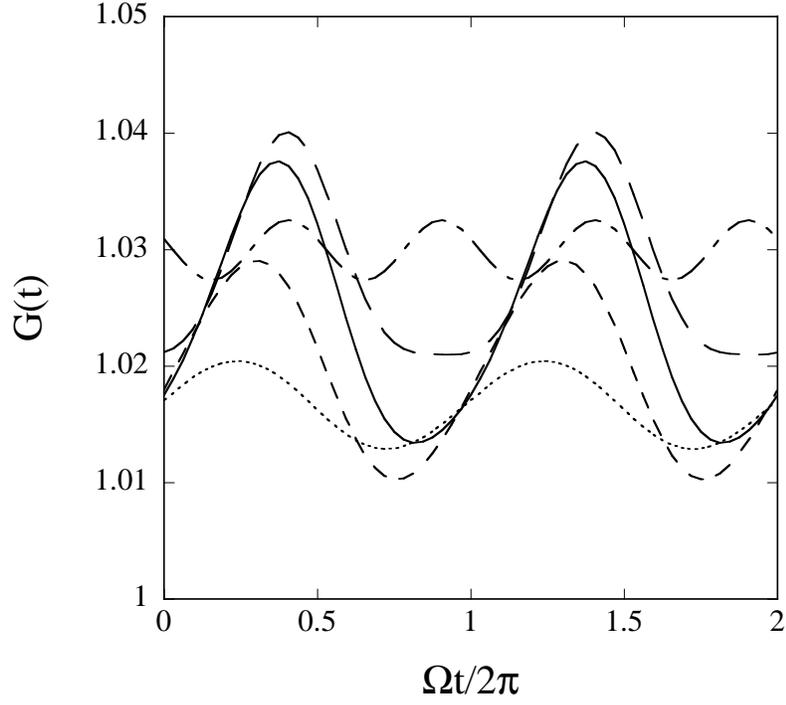,width=0.6\textwidth}
\caption{$G(t)$ as a function of $\Omega t/2\pi$ for the fundamental
odd $r$-mode with $m=1$ for $R=10$km, where $C_1=0.2$, $C_2=0$, $\chi=0$, and
$\bar\Omega=0.2$, and the dotted, short-dashed, solid, long-dashed,
and dash-dotted lines indicate the inclination angle $i=10^\circ$,
$30^\circ$, $50^\circ$, $70^\circ$, and $90^\circ$, respectively.
Here, the amplitude normalization for the mode is given by
$|{\rm Re}[iT_{l_1}(R)/R]|=1$ at the surface.}
\end{figure}

\begin{figure}
\centering
\epsfig{file=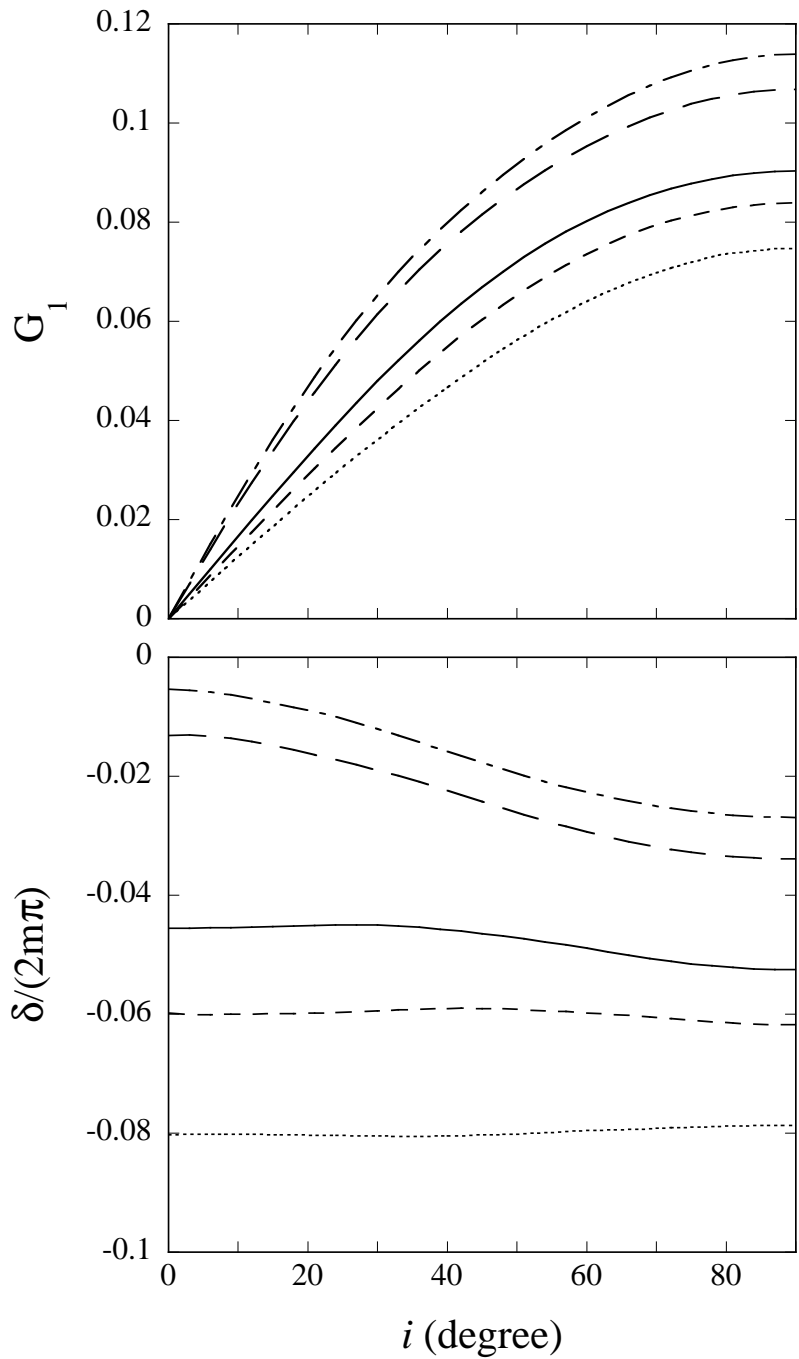,width=0.6\textwidth}
\caption{Same as Figure 3 but for $\bar\Omega=0.4$.}
\end{figure}

\begin{figure}
\centering
\epsfig{file=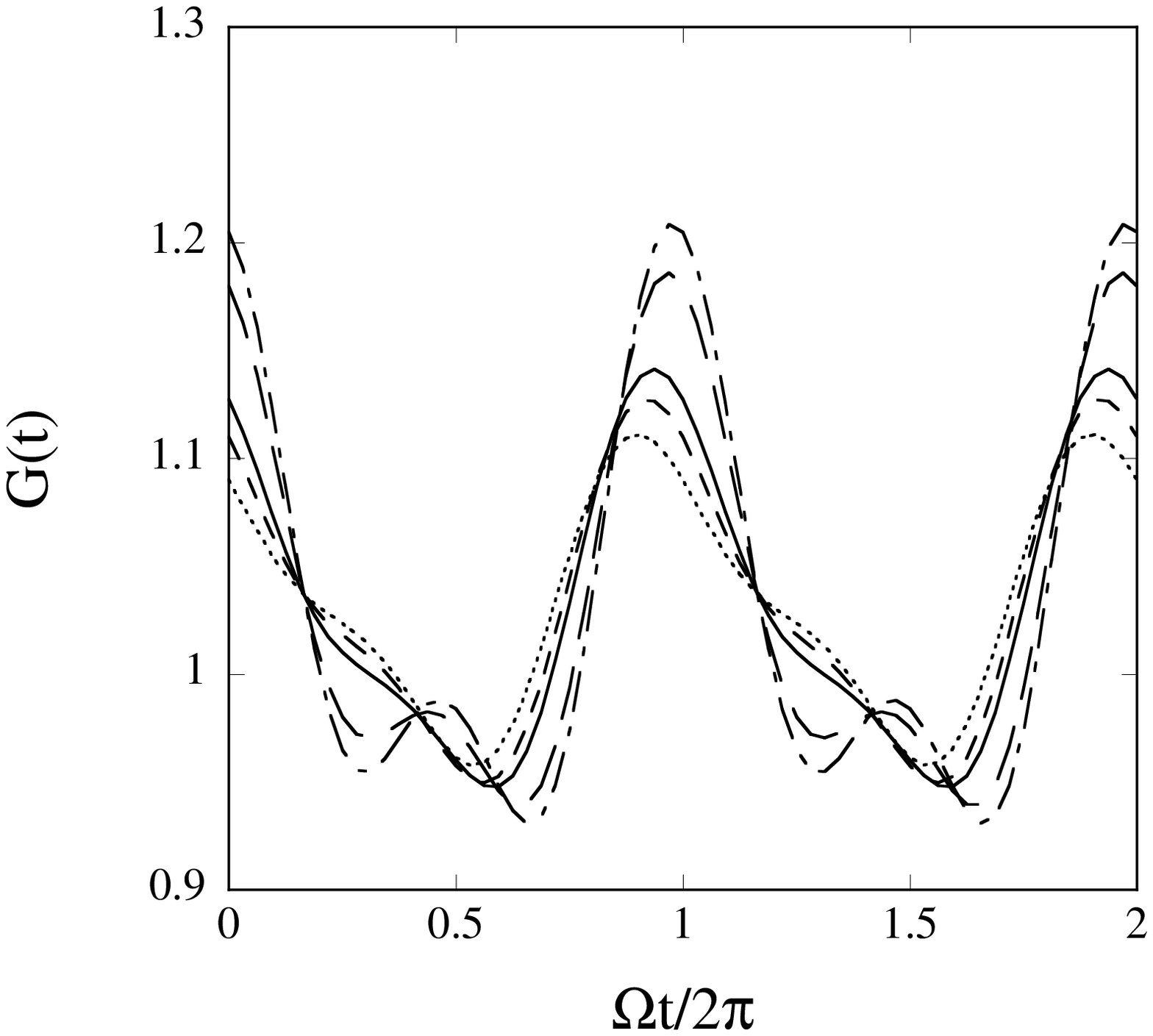,width=0.6\textwidth}
\caption{$G(t)$ as a function of $\Omega t/2\pi$ for the fundamental
even $r$-modes with $m=1$ and $m=2$ for $i=90^\circ$, where
$C_1=C_2=0.2$, $\chi=0$, and $\bar\Omega=0.2$, and 
the various curves represent different radii as in Figure 3.  Here, the amplitude
normalization for the mode is given by $|{\rm Re}[iT_{l_1}(R)/R]|=1$ at the
surface.}
\end{figure}

\begin{figure}
\centering
\epsfig{file=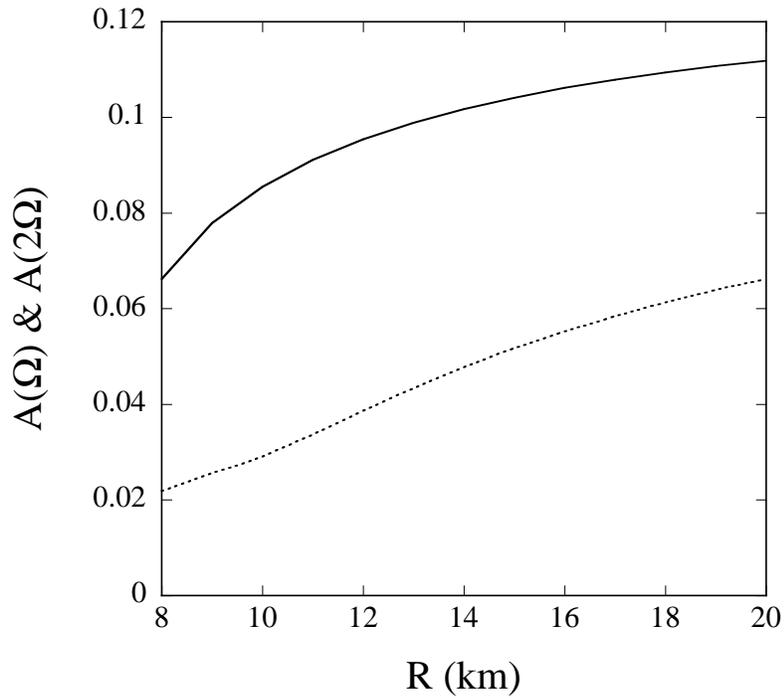,width=0.6\textwidth}
\caption{Fourier amplitudes $A(\Omega)$ and $A(2\Omega)$ as a function
of the radius $R$ for the fundamental even $r$-modes with $m=1$ and
$m=2$ for $i=90^\circ$, where $C_1=C_2=0.2$, $\chi=0$, and
$\bar\Omega=0.2$, and the solid and dotted lines indicate $A(\Omega)$
and $A(2\Omega)$, respectively.  Here, the amplitude normalization for
the modes is given by $|{\rm Re}[iT_{l_1}(R)/R]|=1$ at the surface.}
\end{figure}

\begin{figure}
\centering
\epsfig{file=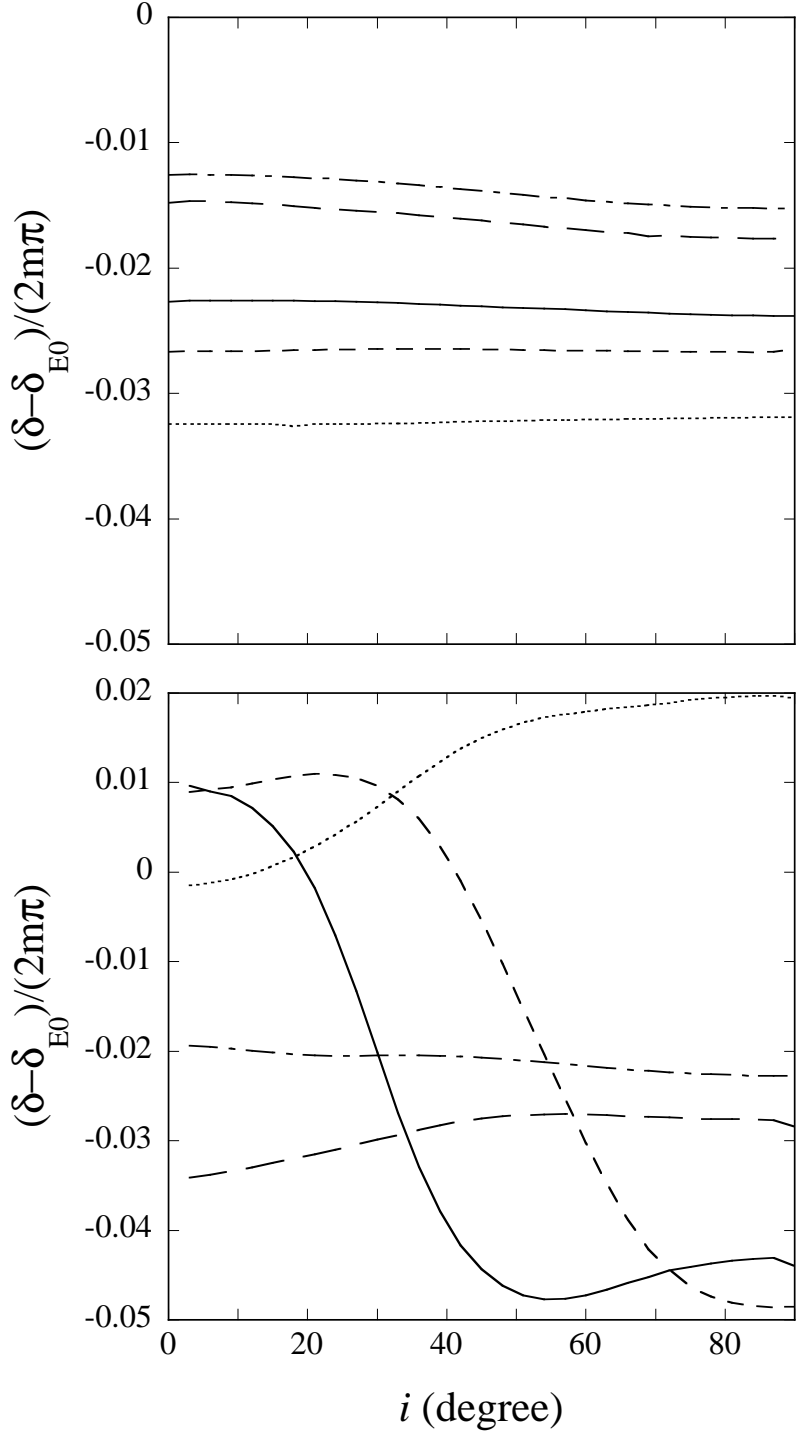,width=0.6\textwidth}
\caption{Phase shift difference $(\delta-\delta_{E_0})/(2m\pi)$ as a function of
the inclination angle $i$ for the fundamental even $m=1$ $r$-mode (upper panel) 
and for the fundamental odd $m=1$ $r$-mode (lower panel), 
where $\bar\Omega=0.2$, $C_1=0.2$, $C_2=0$, and $\chi=0$, and 
the various curves represent different radii as in Figure 3.  
Here, we have assumed $k_BT_0=2.3$kev
and $E_\infty=E_0=1$kev to calculate the function $G_{E_\infty}(t)$ and the phase
shift $\delta_{E_0}$. Negative values of
$(\delta-\delta_{E_0})/(2m\pi)$ indicate hard leads.}
\end{figure}

\begin{figure}
\centering
\epsfig{file=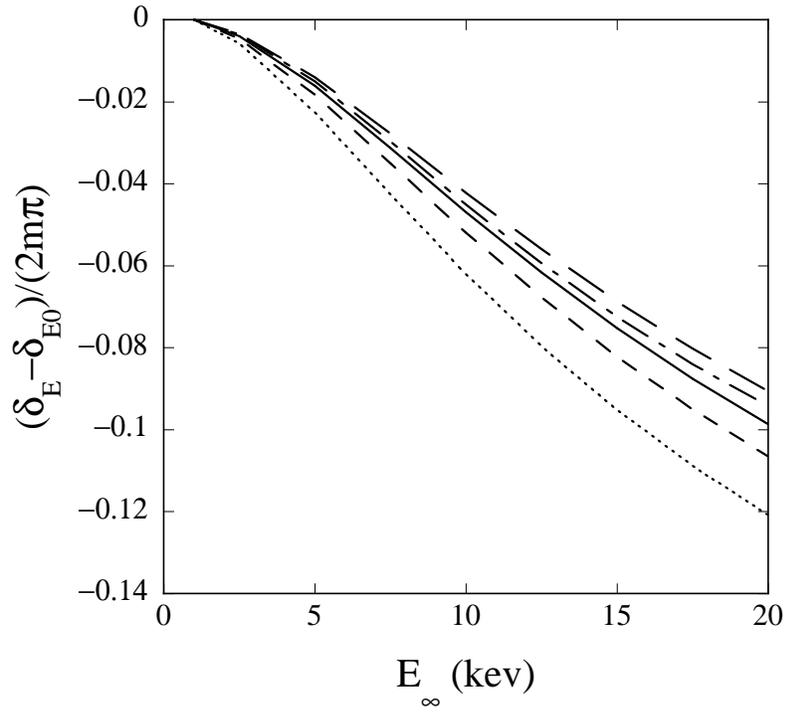,width=0.6\textwidth}
\caption{Phase shift difference $(\delta_E-\delta_{E_0})/(2m\pi)$ as a function of
the photon energy $E_\infty=E$ for the fundamental even $m=1$ $r$-mode, 
where $\Omega_\infty/2\pi=400$Hz, $C_1=0.2$, $C_2=0$, $\chi=0$, and $i=60^\circ$, 
and the various curves represent different radii as in Figure 3.  
Here, we have assumed $k_BT_0=2.3$kev
and $E_\infty=E_0=1$kev to calculate the function $G_{E_\infty}(t)$ and the phase
shift difference $(\delta_E-\delta_{E_0})/(2m\pi)$, the negative values of which
indicate hard leads.}
\end{figure}

\begin{figure}
\centering
\epsfig{file=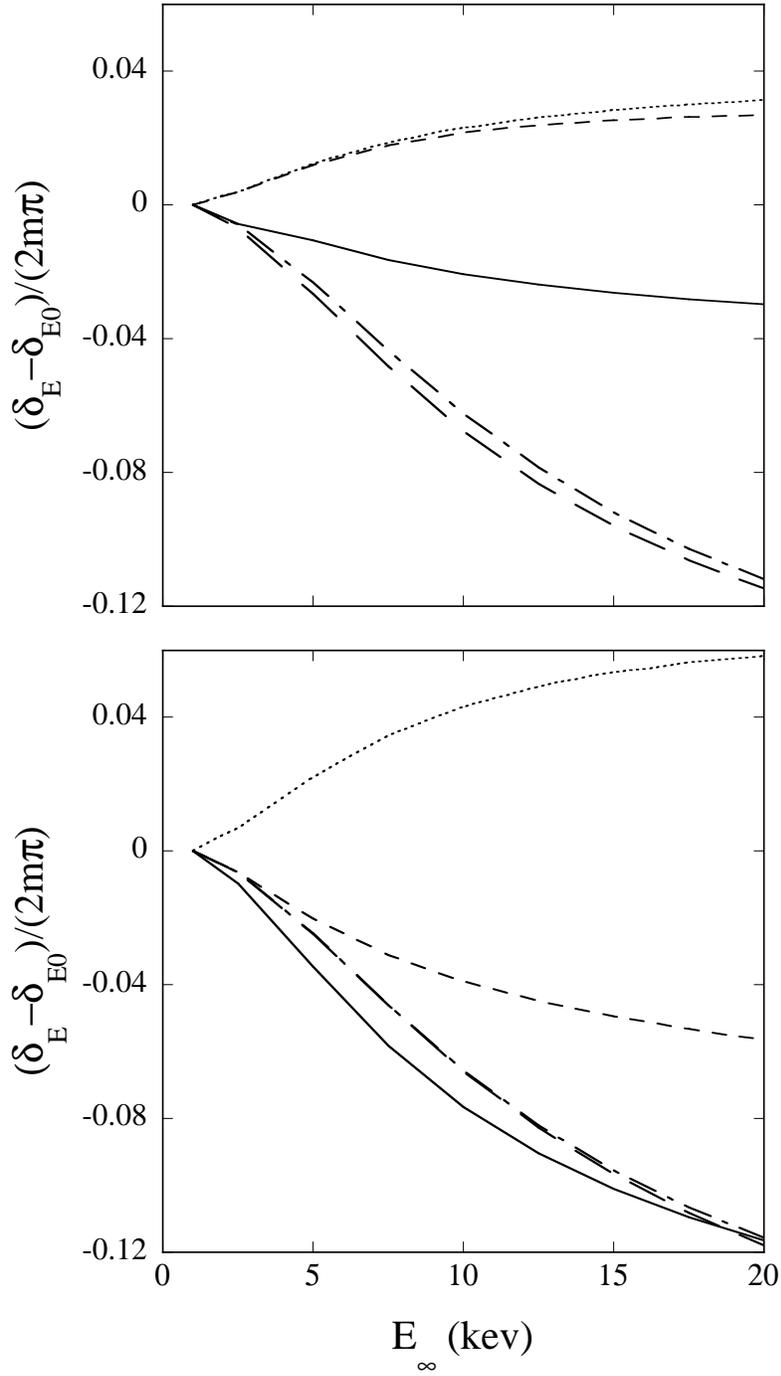,width=0.6\textwidth}
\caption{Phase shift difference $(\delta_E-\delta_{E_0})/(2m\pi)$ as a function of
the photon energy $E_\infty=E$ for the fundamental odd $m=1$ $r$-mode 
for $i=30^\circ$ (upper panel) and for $i=60^\circ$ (lower panel), 
where $\Omega_\infty/2\pi=400$Hz, $C_1=0.2$, $C_2=0$, and $\chi=0$, 
and the various curves represent different radii as in Figure 3.  
Here, we have assumed $k_BT_0=2.3$kev
and $E_\infty=E_0=1$kev to calculate the function $G_{E_\infty}(t)$ and the phase
shift difference $(\delta_E-\delta_{E_0})/(2m\pi)$, the negative values of which
indicate hard leads.}
\end{figure}

\end{document}